\documentclass[%
superscriptaddress,
reprint,
showpacs,
amsmath,
amssymb,
aps,
pra,
]{revtex4-1}

\usepackage{graphicx}
\usepackage{dcolumn}
\usepackage{bm}
\usepackage{amsmath}
\usepackage{amssymb}
\usepackage{xcolor}
\usepackage{multirow}

\usepackage{graphicx,type1cm,eso-pic,color}

\begin{document}

\title{
Electrostatic guiding of the methylidyne radical at cryogenic temperatures
}

\author{David M. Lancaster}
\affiliation{Department of Physics, University of Nevada, Reno NV 89557, USA}
\author{Cameron H. Allen}
\affiliation{Department of Physics, University of Nevada, Reno NV 89557, USA}
\author{Kylan Jersey}
\affiliation{Department of Physics, University of Nevada, Reno NV 89557, USA}
\author{Thomas A. Lancaster}
\affiliation{Department of Physics, University of Nevada, Reno NV 89557, USA}
\affiliation{Department of Biomedical Engineering, University of Michigan, Ann Arbor MI 48109, USA}
\author{Gage Shaw}
\affiliation{Department of Physics, University of Nevada, Reno NV 89557, USA}
\author{Mckenzie J. Taylor}
\affiliation{Department of Physics, University of Nevada, Reno NV 89557, USA}
\author{Di Xiao}
\affiliation{Department of Physics, University of Nevada, Reno NV 89557, USA}
\author{Jonathan D. Weinstein} 
\email{weinstein@physics.unr.edu}
\homepage{http://www.weinsteinlab.org}
\affiliation{Department of Physics, University of Nevada, Reno NV 89557, USA}

\begin{abstract}

We have produced a cryogenic buffer-gas cooled beam of the diatomic molecular radical CH (methylidyne).  
This molecule is of interest for studying cold chemical reactions and fundamental physics measurements. Its light mass and ground-state structure make it a promising candidate for electrostatic guiding and Stark deceleration, which allows for control over its kinetic energy. This control can facilitate studies of reactions with tuneable collision energies and trapping for precise spectroscopic studies.
Here, we have demonstrated electrostatic guiding of CH with fluxes up to $10^{9}$ molecules per steradian per pulse. 
\end{abstract}




\maketitle

\section{Introduction} \label{intro}

The diatomic molecular radical CH (methylidyne) is of great interest for cold molecule experiments for many reasons. Due to the importance of the CH molecule in astronomical spectra, its ground-state structure is well understood \cite{phelps1966experimental, mccarthy2006detection}. Also, precision spectroscopy of the CH rotational structure can be used to test fundamental physics: by comparing spectra of cold CH molecules on earth to astrophysical spectra from distant sources, one can make sensitive tests of time variation of the fundamental constants \cite{truppe2013search}.

Methylidyne's combination of light mass, moderate electric dipole moment, and the lambda-doublet structure of its $X ^2\Pi_{1/2}$ ground state make it an attractive candidate for Stark deceleration \cite{fabrikant2014method, guang2008new}. Additionally, the level structure of CH may make it a viable candidate for laser cooling to ultracold temperatures \cite{DanielMcCarronDAMOP2019}.

CH is the simplest hydrocarbon, and understanding how it reacts with other neutral and ionic molecules is important for multiple fields \cite{Softley09_UltracoldChemistryReview, brownsword1997kinetics, maksyutenko2011crossed}:
it is an important reaction intermediate in combustion chemistry \cite{miller1990chemical} and an important reactant in astrochemistry \cite{canosa1997reactions, prasad1980model, millar1991interstellar}. 
Due to its simplicity, measurements of reactions with CH are useful to help constrain {\textit {ab initio}} theory \cite{brownsword1997kinetics, fabrikant2014method}. By measuring chemical reactions with CH at cold temperatures, more stringent limits can be placed on theoretical models and an improved understanding of the underlying physics and reaction mechanisms can be obtained \cite{Henson234, greenberg2018quantum, fabrikant2014method, chen2019isotope}.

To be able to study cold chemical reactions, it is important to generate large fluxes of cold molecules. Cryogenic buffer-gas beam (CBGB) sources are one of the highest-flux sources for cold molecular radicals \cite{maxwell2005high}.
For experiments seeking to explore chemistry between CH molecules and trapped ions, one also desires a velocity-tunable beam to measure reaction rates as a function of collision energy  \cite{fabrikant2014method,Henson234}. This can be achieved with a Stark decelerator \cite{PhysRevLett.83.1558}, which can decelerate a buffer-gas beam. However, it is critical to guide the molecules from the CBGB source to the entrance of the decelerator to preserve the phase-space density of the beam \cite{fabrikant2014method}. This guiding is also advantageous for other experiments (such as laser cooling and trapping \cite{barry2014magneto} or precision measurements \cite{andreev2018improved}) that require spatial separation between the CBGB apparatus and the experimental measurement region, so as to work in a location of ultrahigh vacuum or large optical access. 

Electrostatic guiding has been demonstrated for supersonic beams of stable molecules and beams of molecular radicals, and has become an important tool in supersonic beam experiments \cite{van2012manipulation}.
Electrostatic and magnetostatic guiding of cold molecules from a CBGB source has  been demonstrated for chemically \textit{stable} species  \cite{patterson2007bright, van2009electrostatic, patterson2009intense}. The objective of the experiment described here is to create a high-flux cryogenic beam source of methylidyne \textit{radicals}, and guide it using an electrostatic guide. The ultimate goal of this work is to guide a beam of radicals into the entrance of a travelling-wave Stark decelerator as described in Ref. \cite{fabrikant2014method}. The output of the decelerator will then be directed to an ion trap to measure cold chemical reactions between CH molecules and molecular ions \cite{staanum2008probing, willitsch2008cold, greenberg2018quantum, chen2019isotope}. 

This paper describes three different elements of the experiment:
first, testing CH production from numerous materials at 300~K; second, creating and measuring a cryogenic molecular beam using two different cell geometries; and third, electrostatically guiding the radicals with a hexapole.
Because of the unusual behavior of the CH molecules,
we also produced and measured a cryogenic atomic titanium beam for comparison.
The titanium data is presented in Appendix \ref{sect:TiResults}.

\section{Production tests at 300~K} \label{sect:300K}
 
We began the process of optimizing the production of CH molecules in a room-temperature cell. This allowed us to rapidly try many different ablation targets and production methods without having to warm up and cool down a cryogenic system, which can take considerable time. 

The room-temperature cell is made from a  standard KF-40 six-way cross with internal volume of $\sim$65~cm$^3$, where the ablation target is mounted so that it protrudes slightly into the central region from the bottom arm. A 532~nm Nd:YAG laser with 5~ns pulses is used to ablate the target in a helium buffer gas environment, with or without H$_2$ gas present, to create the CH molecules. We measured the production of CH from ablating various target materials without H$_2$ present, including: polystyrene, black and white UHMW polyethylene, polypropylene, paraffin wax, black and white nylon, and polyimide film (i.e. Kapton film). We also measured CH production from four graphite targets by ablating them into a mixture of helium and hydrogen gases. The former method --- ablating a compound into an inert gas --- has traditionally been used for production of molecular radicals in cryogenic environments \cite{weinstein1998spectroscopy, CaHnature}. The latter method ---  ablating a pure element that contains one of the atoms of a diatomic radical into a gas containing the other --- has been demonstrated more recently to provide larger numbers of molecules and, equally importantly, more reliable production \cite{ImperialCollege2017}.
	
To create CH molecules, the ablation laser is focused onto the target with a  175~mm focal length plano-convex lens. CH production showed weak sensitivity to the ablation spot size: moving the lens $\pm$7~mm from the central focus position had little effect on  production. All measurements presented here use ablation energies between 10 and 90 mJ/pulse. The graphite targets have a production threshold around 10~mJ, with peak production occurring at $\sim80$~mJ. All other targets did not produce a measurable signal below 20~mJ/pulse; these targets typically produce peak signals at $\sim 70$~mJ.
 
\begin{figure}[h!t]
    			\begin{center}
     			 \includegraphics[width=\linewidth]{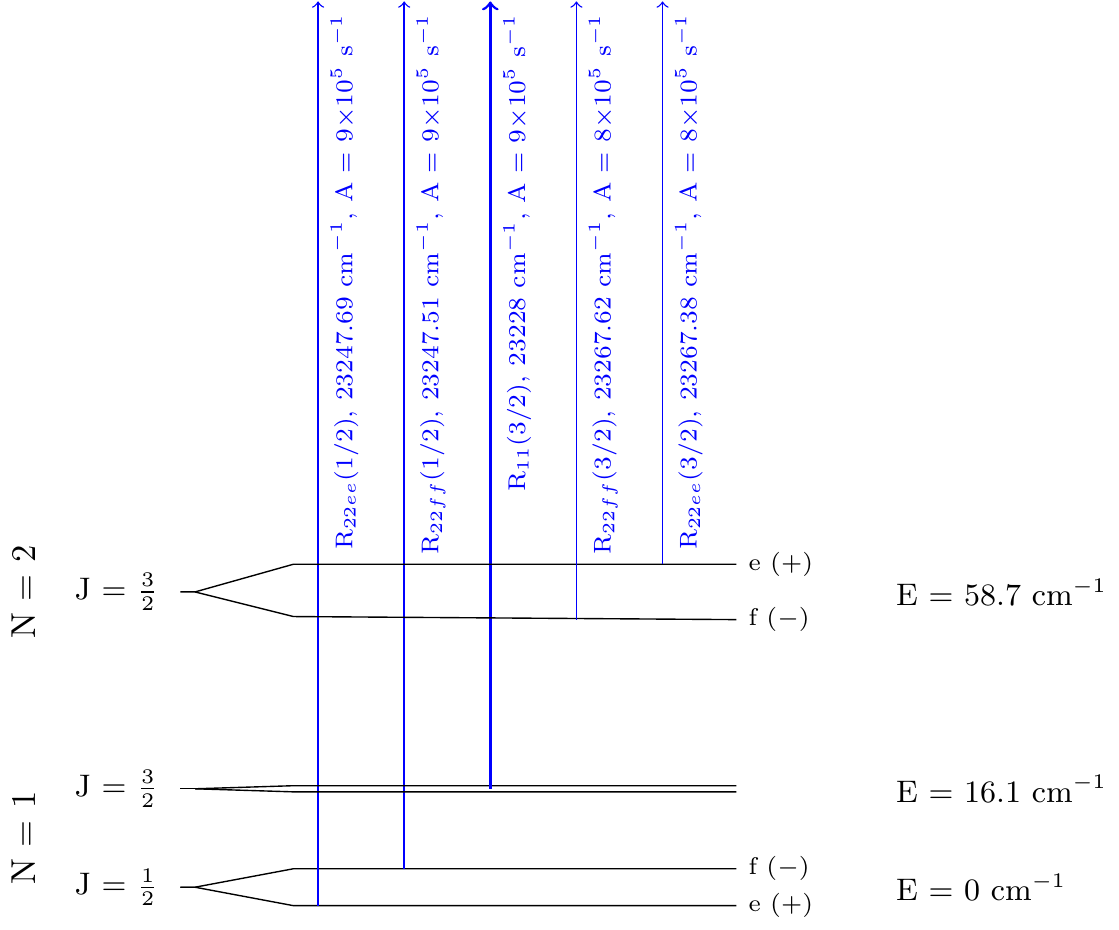}
    					\caption{
    						\label{fig:energy_diagram} 
    						Diagram of the $X^2 \Pi$(v = 0) level of CH showing the $X\leftrightarrow A$ transitions we probed. Each transition is labeled by its name, its transition wavenumber in cm$^{-1}$, and its Einstein $A$ coefficient \cite{CH_wavenum__ZACHWIEJA1995285,CH_microwave_spectroscopy,CH_transition_probabilities}. The naming convention 
    		is from				
				Ref. \cite{CH_wavenum__ZACHWIEJA1995285}. The figure is not drawn to scale; rotational energies are given at the right in wavenumbers. 
					}     
    			\end{center}
		\end{figure}

The production of CH in the room-temperature gas cell is evaluated using direct absorption spectroscopy.  The level structure of the ground state of CH is shown in Fig. \ref{fig:energy_diagram}, along with the optical transitions used in this work for detection. We use a tunable 430~nm external cavity diode laser to produce the probe light.
In the room-temperature tests, we detected the gas-phase methylidyne radicals via triple-pass absorption spectroscopy on the R$_{11}$(3/2) transition at 23228~cm$^{-1}$ \cite{CH_wavenum__ZACHWIEJA1995285} (The $\lambda$-doublet is unresolved in our experiment.) The laser power after passing through the cell is measured using a photodiode and recorded by an oscilloscope. 

From the measured absorption signals, we determine the number of CH molecules produced per ablation pulse. The molecule numbers are calculated assuming a 300~K Doppler width, and assuming the CH density is uniform in the central volume of the KF cross and zero in the arms \cite{BudkerBook,CH_transition_probabilities}. The measured production numbers are listed in Table \ref{Tab:CH_comparison}. We fit the decay of the absorption signal as a function of time to an exponential to give the lifetimes of the CH molecules in the cell, which are also reported in  Table \ref{Tab:CH_comparison}.

	\begin{table}[ht] 
	
\caption{CH production methods with corresponding lifetimes and numbers of molecules produced per ablation pulse \cite{Lancaster2018Thesis}. 
Helium buffer gas densities ranged from $4\times 10^{15}$ to $2 \times 10^{17}$~cm$^{-3}$. Typical densities that gave good production were $2 \times 10^{16}$~cm$^{-3}$ for hydrocarbon targets, and $1 \times 10^{16}$~cm$^{-3}$ for graphite targets. 
Lifetimes cited are those measured at conditions that gave peak absorption signals.
The number of molecules reported here is the number in the 
$N=1, J=3/2$ state; the total number produced is higher.}
\centering

\begin{tabular}{||c|c|c|c||}
\hline
Gas & Ablation target & \begin{tabular}[c]{@{}c@{}}Lifetime \\ ($\mu$s)\end{tabular} & \begin{tabular}[c]{@{}c@{}}Molecules\\ produced \\ ($ \times 10^{11} $)\end{tabular} \\ \hline\hline
\multirow{8}{*}{He} & Paraffin wax & 160 & 8.5  \\ \cline{2-4} 
 & Kapton sheet & 140 & 4.4  \\ \cline{2-4} 
 & White polyethylene & 140 & 4.4 \\ \cline{2-4} 
 & White nylon & 110 & 4.7 \\ \cline{2-4} 
 & Polypropylene & 50 & 4.1 \\ \cline{2-4} 
 & Black polyethylene & 100 & 3.1 \\ \cline{2-4} 
 & Black nylon & 115 & 4.1 \\ \cline{2-4} 
 & Polystyrene & N/A & $ \lesssim 1 $ \\ \hline
\multirow{3}{*}{H2} & Compressible graphite & 250 & 21 \\ \cline{2-4} 
 & Conductive graphite & 250 & 24 \\ \cline{2-4} 
 & UFGC graphite & 225 & 26 \\ \hline
\multirow{3}{*}{\begin{tabular}[c]{@{}c@{}}94\% He\\ +\\ 6\% H2\end{tabular}} & UFGC graphite & 240 & 22 \\ \cline{2-4}
 & Compressible graphite & 210 & 22 \\ \cline{2-4}
 & Pyrolytic graphite & 240 & 23 \\ \hline
\end{tabular} 
 \label{Tab:CH_comparison}

\end{table}

The best hydrocarbon target in terms of both production and cell lifetime was paraffin wax, producing nearly $10^{12}$ molecules (in the $N=1, J=3/2$ state) per pulse. The graphite targets showed better performance than the hydrocarbon targets, with roughly twice as many molecules produced and cell lifetimes $\sim 50\%$ longer. Moreover, the CH production from the graphite targets was more consistent shot-to-shot. Consequently, we chose graphite ablation into a mixture of hydrogen and inert gases as our method of molecule production for our cryogenic cell. The majority of data obtained in the cryogenic experiment were for this production method.
In addition to the species listed in Table \ref{Tab:CH_comparison}, we attempted to create targets using iodoform powder.
Unfortunately, we were unable to do so, as described in Appendix \ref{sect:iodoformtgt}. 

\section{Cryogenic measurements: standard cell} \label{sect:standard_cell}
	
The cryogenic apparatus used has been described previously \cite{Nozzles2018,Lancaster2018Thesis,MackTaylorThesis2018}. Briefly, the apparatus consists of a rectangular cuboid vacuum chamber made of aluminum with a 50~K inner radiation shield and houses a pulsed-tube refrigerator (Cryomech PT415).  Inside the shield are two cryopumps and an ablation cell, all made from 110 (ETP) copper. The cell is thermally connected to the second stage of the refrigerator and has an internal volume of $\sim24$~cm$^3$. The ablation cell is typically held at 16~K, and the cryopumps at around 4~K. This cryopump temperature allows us to pump both Ne and H$_2$ gases.

The buffer-gas beam is extracted from the cell using simple cylindrical hole apertures, with the most common being 3/16'' in diameter. Windows are on the vacuum chamber and ablation cell such that we can simultaneously probe absorption inside the cell and either absorption or fluorescence of the extracted beam, about 6~cm after the cell exit. 

				\begin{figure}[h!t]
    			\begin{center}
     			 \includegraphics[width=\linewidth]{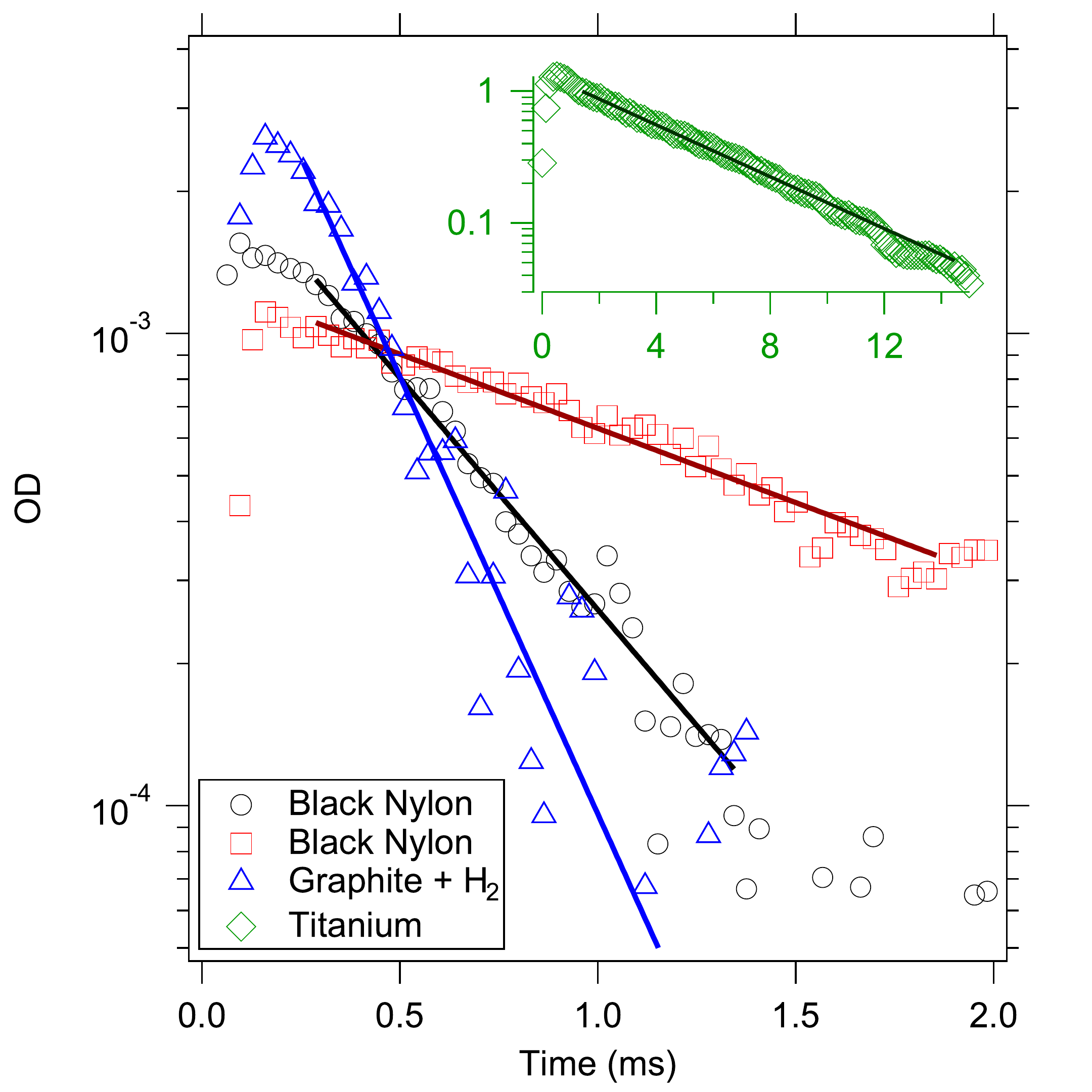}
    					\caption{
    						\label{fig:Target_tau_compar} 
    						Lifetime comparison  in the ``standard cell''. Shown are examples from: graphite ablation into H$_2$ and Ne (triangles); black nylon ablation into Ne (squares and circles), with both examples being from the same target; and Ti ablation into Ne (diamonds, inset). Note that the inset graph has the same units, but different scale, as the main graph. 
    						 Lifetimes are as follows -- black nylon: $\tau_{\mathrm{squares}} = 1.4$~ms, $\tau_{\mathrm{circles}} = 0.43$~ms; Graphite + H$_2$: $\tau_{\mathrm{triangles}} = 0.23$~ms; Titanium: $\tau_{\mathrm{diamonds}} = 4.4$~ms.
					}     
    			\end{center}
		\end{figure}

We produce CH in this cell by ablation of graphite targets into mixtures of neon and hydrogen gas, and by ablation of black nylon into neon gas. For ablation, we use 
a frequency-doubled Nd:YAG laser operating at 10~Hz with energies of 10--80~mJ~per~pulse. The ground rotational state of CH was detected by single-pass absorption spectroscopy of the R$_{22}(1/2)$ transitions \cite{CH_wavenum__ZACHWIEJA1995285}, as shown in Fig. \ref{fig:energy_diagram}. The maximum optical depth, OD, observed in the cell was on the order of 1$\times 10 ^{-2}$. Assuming the molecules are evenly distributed through the central region of the cell, this corresponds to  $\sim 6\times 10^{10}$ molecules in a single $f$ or $e$ state of the ground rovibrational state.

\begin{figure}[h!t]
    			\begin{center}
     			 \includegraphics[width=\linewidth]{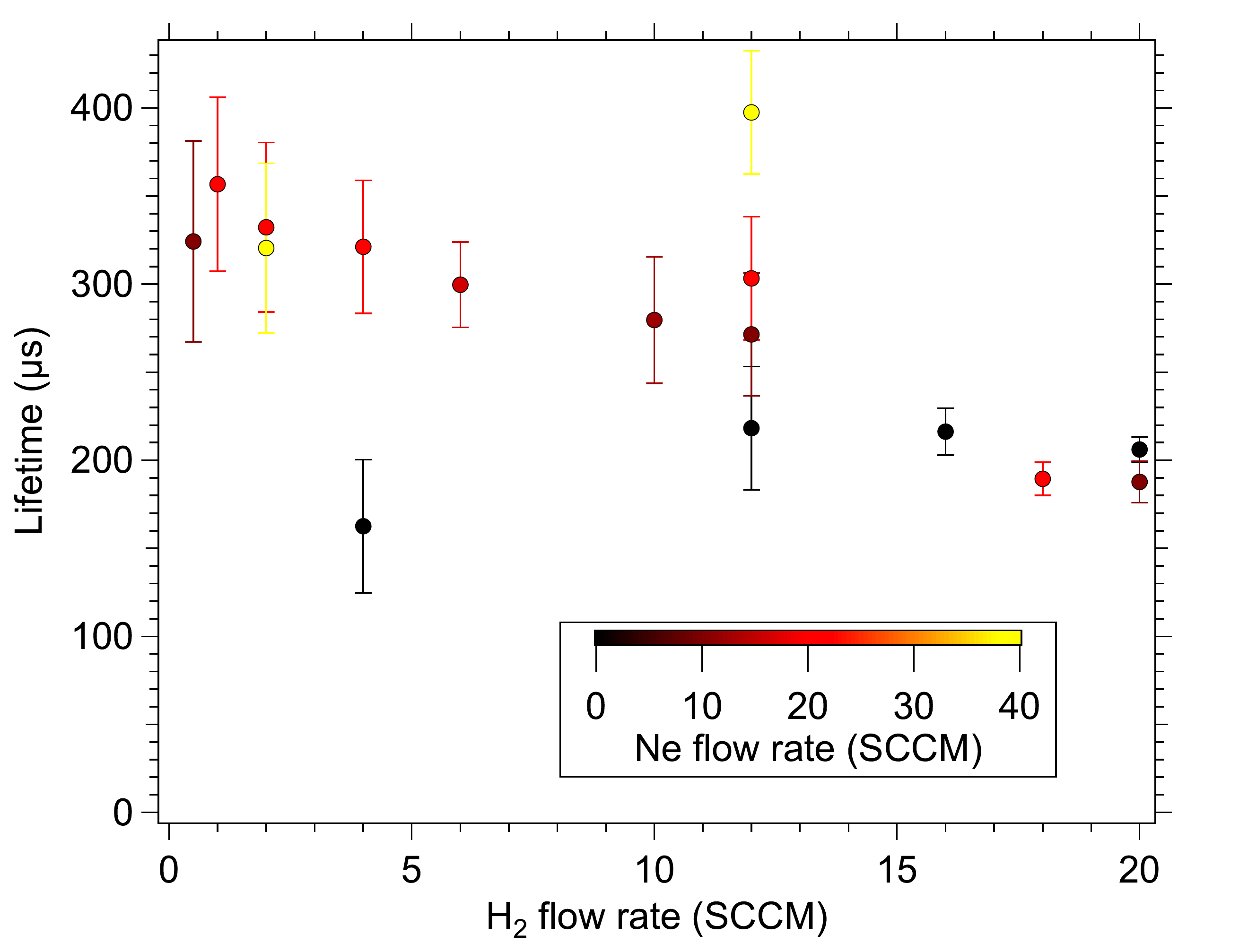}
    					\caption{
    						\label{fig:old_cell_CH_lifetimes} 
    						Lifetimes of CH  produced by ablation of a graphite target into a mixture of hydrogen and neon in the standard cell, as discussed in the text. The error bars are the standard error of the mean.
					}     
    			\end{center}
		\end{figure}

Raw data for in-cell absorption can be seen in Fig. \ref{fig:Target_tau_compar}, with comparison to Ti absorption in the same cell. Unfortunately, the in-cell lifetime for CH production by ablation of graphite is short, and comparable to 300~K observations. The behavior of the lifetime as a function of buffer-gas density and composition is shown in Fig. \ref{fig:old_cell_CH_lifetimes}. With non-zero neon flow, the lifetime is typically observed to decrease as more H$_2$ is introduced. With no neon flow, the lifetime is relatively constant as more H$_2$ is added. 
		
Comparing the CH lifetime behavior in Fig. \ref{fig:old_cell_CH_lifetimes} to extraction and lifetime characteristics for a beam of titanium atoms (detailed in Appendix \ref{subsec:StandardCellTiResults}) would suggest that the dominant loss process is neither diffusion to the cell walls nor extraction from the cell through the aperture.

The cell lifetime data would suggest the vast majority of CH is destroyed before it can be extracted into the buffer-gas beam. Consistent with this expectation, we detect no fluorescence in the beam outside the cell to within our noise (OD~$<$~$10^{-4}$). Potential causes of the rapid destruction of CH include chemical reactions, clustering, and adsorption onto dust particles. 
		
Considering chemical reactions with hydrogen, the two-body chemical reaction CH~+~H$_2 \rightarrow$~CH$_2$~+~H is endothermic, requiring 10~kJ/mol to proceed \cite{NISTChemWebBook}, so it should be highly suppressed in the cold environment. However, three-body reactions such as CH+H$_2$+Ne $\rightarrow$ CH$_3$+Ne are energetically allowed \cite{NISTChemWebBook}. For any such reactions, we would expect the lifetime to decrease as the H$_2$ density is increased. While higher hydrogen flow rates do lead to slightly shorter lifetimes, as seen in Fig. \ref{fig:old_cell_CH_lifetimes}, the dependence on hydrogen flow is much weaker than one would expect if this was the dominant loss process. In fact, when no neon is present, increasing the H$_2$ density has almost no effect on lifetime.
		
Clustering with the neon buffer gas (via three-body collisions \cite{quiros2017cold}) could be a loss mechanism for CH molecules in the cell. However, replacing neon with helium yielded similar lifetimes. Additionally, the lifetime increases with increasing neon flow (for fixed H$_2$ flow). This is not what one would expect for clustering with the buffer gas, where adding more neon should reduce the lifetime. 
		
Finally, the most plausible reason that the CH lifetime is short is due to other species produced by the ablation: either unidentified chemical species that react with CH or dust. If the ablation produces a large amount of dust particles distributed throughout the cell, CH would be expected to adsorb onto the surface of those particles. This mechanism has been posited previously to explain short molecule lifetimes in cryogenic experiments involving ablation of non-metallic targets \cite{weinstein1998spectroscopy}. This mechanism is supported by the longer lifetimes seen in subsequent experiments in the same cell employing a nylon ablation target, as seen in Fig. \ref{fig:Target_tau_compar}. Unfortunately, we have only limited data for the cryogenic cell with this ablation target, but nylon has demonstrated significantly longer cell lifetimes than was observed for graphite, indicating that it is some other species (such as dust) produced by graphite ablation that is limiting the cell lifetime. As seen in Fig. \ref{fig:Target_tau_compar}, ablating the nylon target produces CH with widely varying cell lifetimes shot-to-shot. We suspect that not only is nylon somewhat inconsistent in its production of CH, but that it is also inconsistent in its production of dust.

\section{Cryogenic measurements: Small-volume cell} \label{sect:small_vol_cell}

We attempted to work around the short lifetime of CH molecules in the cell by redesigning the cell for a shorter extraction time. 
To reduce the extraction time, we reduce the volume of our cell. Previous work with helium buffer gas and CaF molecules has shown that internal volumes of $\sim$3~cm$^3$ can produce a cold beam of molecules with a cell extraction times on the order of 100~$\mu$s \cite{ImperialCollege2017}.

\begin{figure}[h!t]
    			\begin{center}
     			 \includegraphics[width=\linewidth]{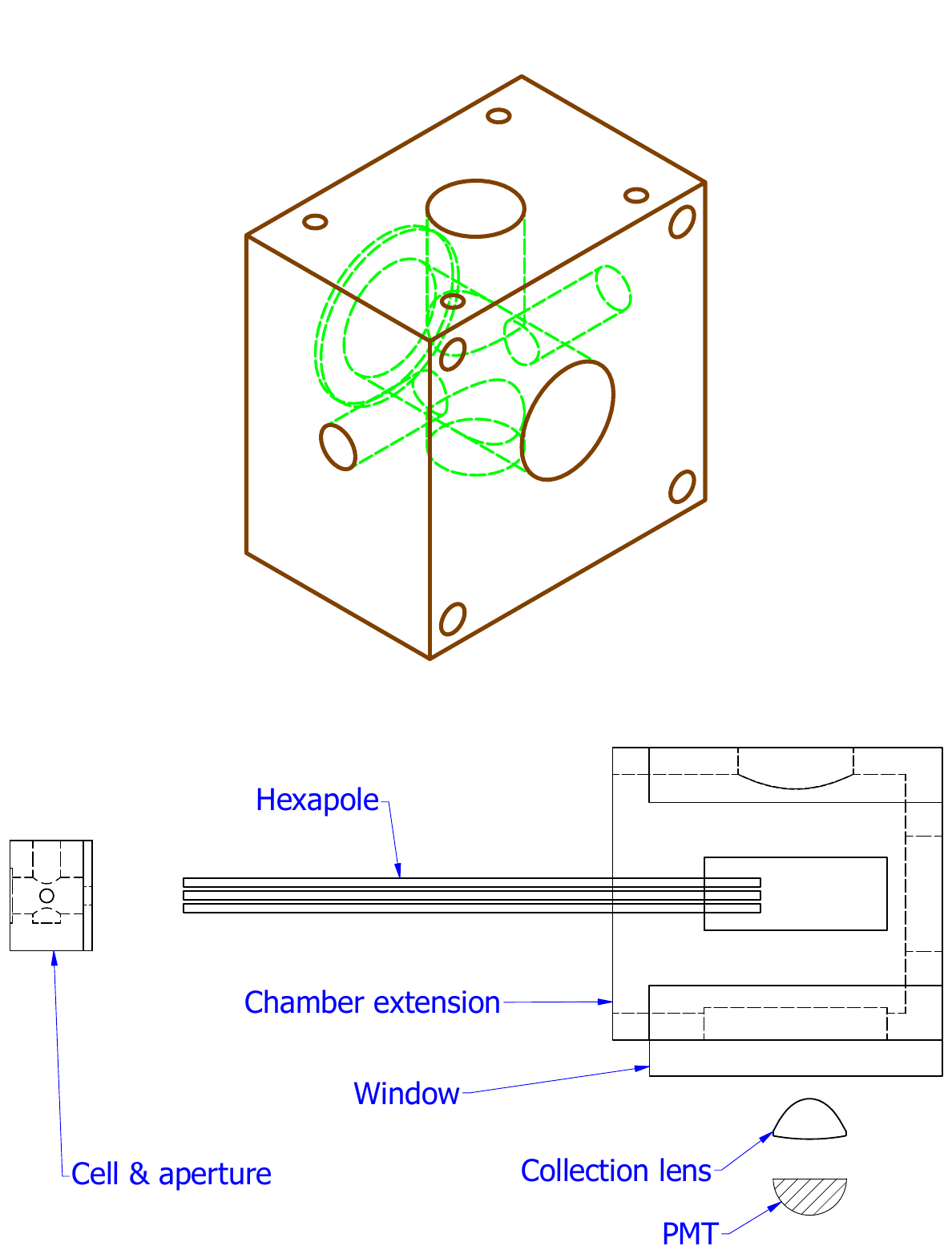}
    					\caption{
    						\label{fig:chamber_schematics} 
    						Top: a wireframe schematic of the small-volume cell, with internal structure shown by dashed green lines. \\	Bottom: a schematic of the hexapole setup. Not shown are the chamber body, which encloses the rest of the hexapole rods and cell, the gas feedthrough for the cell, and the cryopumping surface that the hexapole fits through. All components are drawn to scale, with the cell being 1.5" tall for reference. They are also placed to scale, with the exception of the vertical positioning of the condenser lens and PMT.
					}     
    			\end{center}
		\end{figure}

A schematic of the small-volume cell is shown in Fig. \ref{fig:chamber_schematics}. The internal volume is $\sim 5$~cm$^3$. The main bore has a flared entrance to ensure compatibility with an existing gas inlet \cite{Nozzles2018}. The ablation laser enters through a windowed port on the top of the cell and strikes a cylindrical ablation target that sits in a small cutout underneath the main bore. Two windowless detection ports are used in detecting absorption; some gas will escape through these ports, but due to their small diameter relative to the main bore, we assume this will not significantly alter the gas flow. We use the same style of aperture as the standard cell, but with a larger diameter of 1/4" for faster extraction.

	\subsection{CH production}
	
\begin{figure}[h!t]
    			\begin{center}
     			 \includegraphics[width=\linewidth]{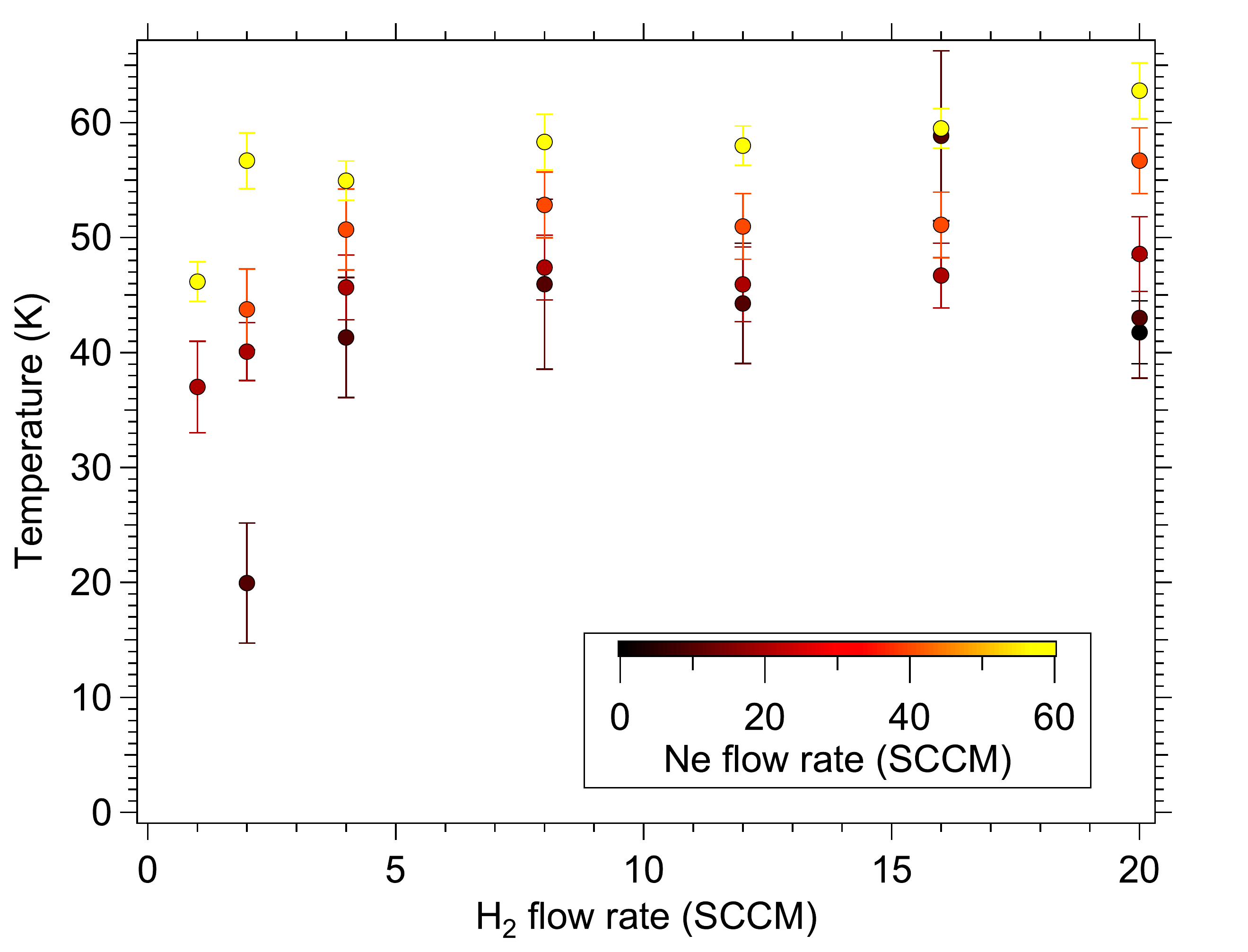}
    					\caption{
    						\label{fig:new_cell_CH_IC_temps} 
    						In cell translational temperature of CH for the small-volume cell geometry calculated from a spectrum taken 30--200~$\mu$s (most of the pulse) after ablation.
					}     
    			\end{center}
		\end{figure}

In-cell absorption measurements provide the translational temperature of the CH molecules, as shown in Fig. \ref{fig:new_cell_CH_IC_temps}. While these temperatures are significantly higher than the cell wall temperatures (measured prior to ablation), we note that these temperatures are comparable to titanium atom temperatures (Fig. \ref{fig:new_cell_Ti_temps_IC}) measured at similar times after the ablation pulse, as described further in Appendix \ref{subsec:SmallVolCellTiResults}.

In-cell absorption measurements show CH production comparable to the standard cell, yielding  $2 \times 10^{11}$ molecules in the $N=1,J=1/2$ rotational state. We note that the production shows dependence on Ne flow rate; with no neon flow and 20~SCCM of H$_2$ we observe an OD of $4 \times 10^{-3}$, and adding 60~SCCM of Ne yields an OD of $2.4 \times 10^{-2}$. 	We also measure large numbers of molecules in excited rotational states. We measure the  populations of the  $N=1, J=3/2$ and $N=2, J=1/2$ levels from the absorption signals of the R$_{22}$(3/2) \& R$_{22}$(1/2) transitions. From fitting the relative populations to a Boltzmann distribution, we extract a rotational temperature of $37\pm$7~K. The uncertainty is dominated by the shot-to-shot fluctuations in production.

The measured lifetime in this new geometry is similar to the standard cell, varying from 100--200~$\mu$s depending on experimental conditions. The lifetime does increase with neon flow rate: at a H$_{2}$ flow of 12~SCCM, adding 10~SCCM of neon yields a lifetime of 100~$\mu$s; adding 60~SCCM of neon yields a lifetime of 160~$\mu$s. A 200~$\mu$s lifetime is achieved at 1~SCCM of H$_2$ and 60~SCCM of Ne. The short lifetimes are consistent with the shorter extraction times observed from the small volume cell. As described in Appendix \ref{subsec:SmallVolCellTiResults}, our extraction time for Ti is reduced by over an order of magnitude. These faster extraction times allow us to see CH molecules in the buffer-gas beam.

	\subsection{Small-volume cell beam}
	\label{subsec:SmallCellBeam}

To detect CH in the beam outside the cell, we use laser-induced fluorescence (LIF) spectroscopy. The molecules are excited by a 430~nm tuneable diode laser, operating at typical powers of 1--1.5~mW, with a beam diameter of 5--6~mm. The fluorescence --- in a direction perpendicular to the excitation beam --- is detected by a photomultiplier tube (PMT). We vary the power of the probe laser to ensure that the measurements are not significantly altered by optical pumping effects or saturation of the transition.
  
For calculating the number of molecules extracted, we assume that the pulse of molecules are moving out of the cell with a Gaussian velocity distribution in the radial directions, and uniform in the axial direction \cite{Nozzles2018}. We assume the primary broadening source is Doppler broadening. Finally, we assume that the extracted beam will be large compared to the probe beam at the probing location, 6~cm from the cell exit aperture, as expected from the high transverse temperatures seen with titanium beams (see Appendix \ref{subsec:SmallVolCellTiResults}).

Due to the geometry of our fluorescence detection, we cannot accurately measure the transverse velocity spread of the beam (the detectable transverse spatial extent of the beam is restricted by an aperture). This also prevents us from measuring the total number of molecules extracted from the cell. However, we can measure the flux of molecules per solid angle (on-axis), and find a flux of ~10$^{10}$ molecules~per~steradian~per~pulse at an ablation energy of 50~mJ. Similar to the in-cell production, we observe that the number of molecules per steradian of CH increases as the neon flow rate increases. At a H$_2$ flow of 20~SCCM, we observe a flux of $\sim 3 \times 10^9$~molecules per steradian per pulse. If we add 60~SCCM of neon we observe $\sim 2.5 \times 10^{10}$~molecules per steradian per pulse. This increase is higher than expected given the in-cell production; we assume the additional increase is due to better extraction.
	
 Axial beam characteristics are determined from LIF spectra collected with the excitation laser running counter to the propagation of the molecular beam; forward velocities are calculated using Doppler shifts, and velocity spreads from Doppler broadening. We compare these results to a beam of titanium atoms, detailed in Appendix \ref{subsec:SmallVolCellTiResults}.
	
\begin{figure}[h!t]
    			\begin{center}
     			 \includegraphics[width=\linewidth]{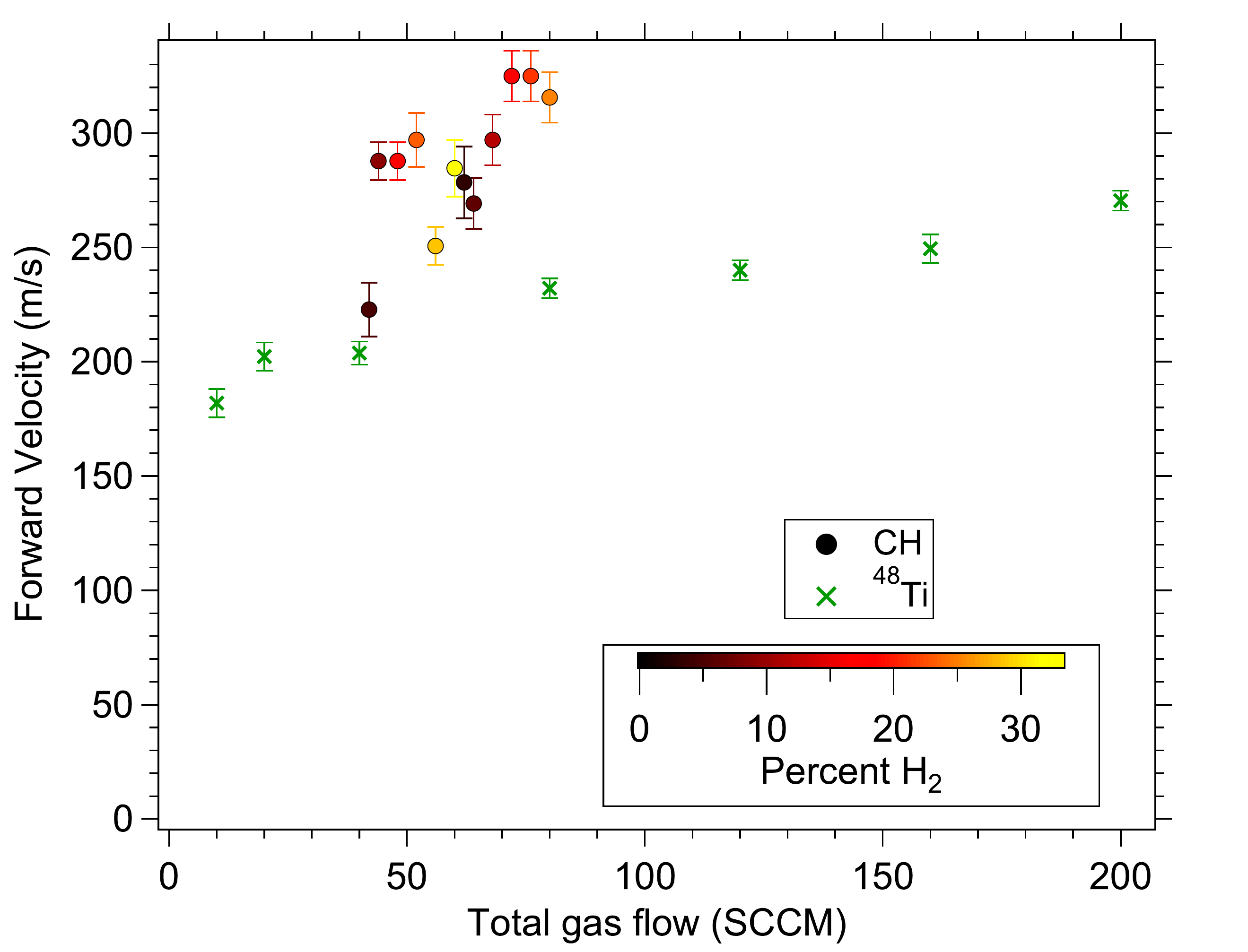}
    					\caption{
    						\label{fig:new_cell_Forward_velocities} 
    						Forward velocity for both Ti and CH beams. Note that the x-axis is the total gas flow, and the color denotes the percent of H$_2$ composition of the total gas.
					}     
    			\end{center}
		\end{figure}

The forward velocity results are shown in Fig. \ref{fig:new_cell_Forward_velocities}. We observe higher forward velocities for CH than what we measured for the titanium beam. We suspect this is partially due to the earlier times of CH extraction (due to the shorter in-cell lifetime), which correspond to higher in-cell gas temperatures. It may also be partially due to the inclusion of hydrogen gas, which has a higher velocity (although we see no strong dependence with H$_2$ fraction) and partially due to the lighter mass of CH relative to titanium.

				\begin{figure}[h!t]
    			\begin{center}
     			 \includegraphics[width=\linewidth]{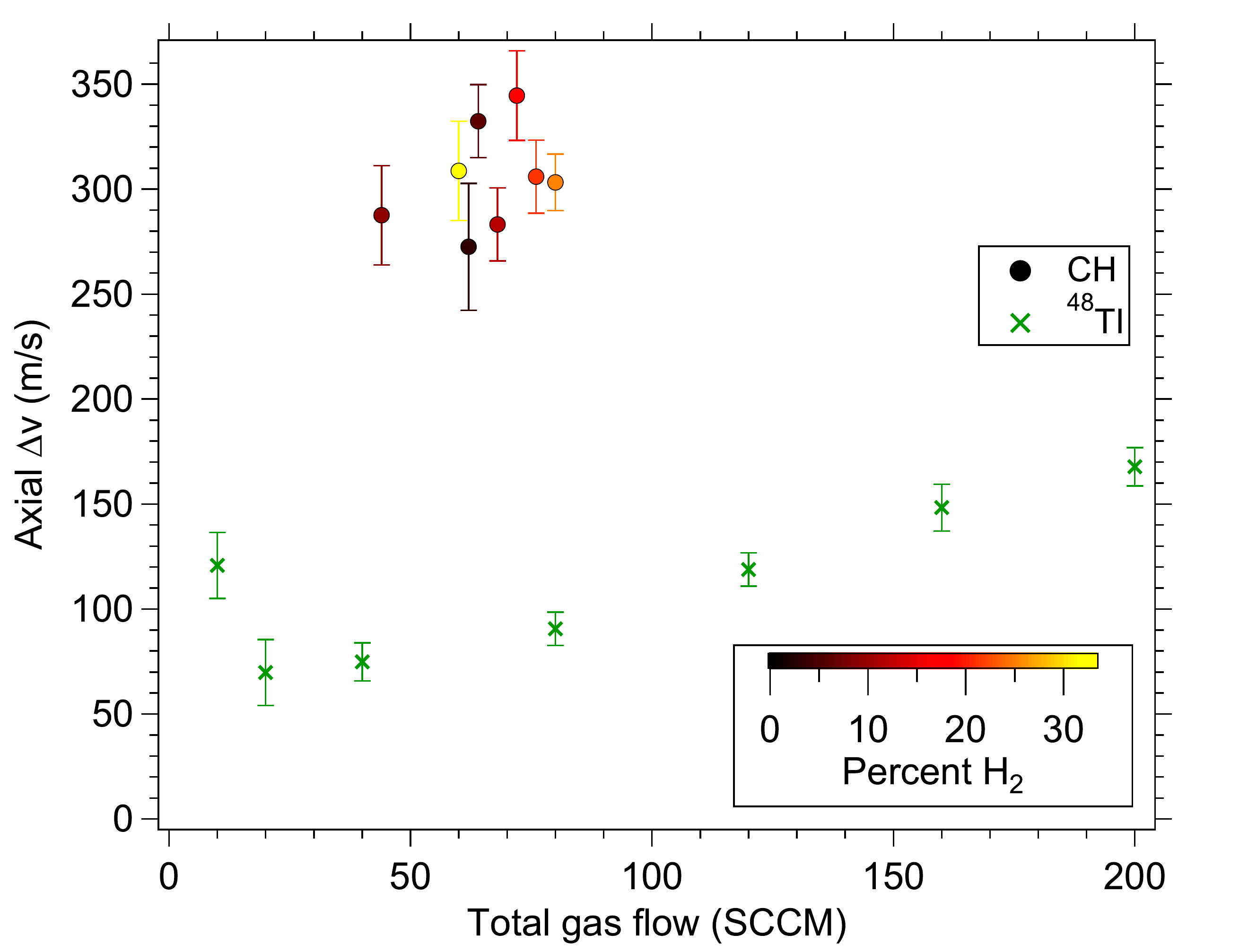}
    					\caption{
    						\label{fig:new_cell_axial_temps} 
    						Axial velocity spread FWHM for both Ti and CH beams from the small-volume cell. This is also plotted vs total gas flow and colored according to H$_2$ composition. Interestingly, the titanium beam seems to be exhibiting expansion cooling at low flow rates, and substantial heating at high flow rates. 
					}     
    			\end{center}
		\end{figure}

The axial velocity spread of the CH beam, as shown in Fig. \ref{fig:new_cell_axial_temps}, is substantially larger than Ti. We believe this is due both to the lighter mass of CH and to the earlier times at which the molecules exit the cell. The velocity spreads shown correspond to temperatures ranging from 20--35~K. The rotational temperature of the beam is measured to be $44\pm5$~K.

We note that these temperatures are significantly higher than previously seen in cryogenic buffer-gas beams produced by small-volume cells \cite{ImperialCollege2017}. We believe the higher temperatures are partially due to the earlier extraction times (see Appendix \ref{sect:TiResults}), but this is not fully understood. These high temperatures are not favorable for guiding. However, our molecular flux is sufficiently large that we expected to be able to observe guided CH, and thus we proceeded to test guiding using this CH beam.

\section{Cryogenic measurements: guiding} \label{sect:guiding}

The experimental setup was modified to accommodate the electrostatic guide, and the probe location for the beam was moved to after the guide. We used a 200~mm long hexapole electrostatic guide with rod radii of 1.5~mm and a bore diameter of 5~mm to guide our molecules. To fit this into our chamber, we added a small extension to the chamber with windows to allow probing at different locations after the hexapole exit. A schematic of the guiding setup is shown in Fig. \ref{fig:chamber_schematics}. We placed our hexapole entrance roughly 30~mm away from the cell aperture, and we probed the guided beam 25~mm after the hexapole's exit. This is sufficiently far from the electrodes that Stark shifts of the transitions should be negligible.

We performed LIF with a laser beam waist of $\sim 9$~mm and typical powers of 1.2~mW. Though this would seem to be a high intensity for the relatively weak optical transitions of CH, we did not observe any saturation of signal with laser power. Due to the low transverse velocity spread of guided molecules, we expect that the waist of our LIF beam is significantly wider than the beam of guided molecules at our measurement location.
We convert our LIF measurements to OD via calculations of collection efficiencies and detector properties.
Unlike the measurements made in the main chamber, the OD is too low to use absorption measurements to directly calibrate the LIF signal.
We note that for the calibrated OD measurements in the main chamber,  our LIF calculations underestimated the true OD by a factor of 3; we expect similar accuracy here.

\begin{figure}[h!t]
    			\begin{center}
     			 \includegraphics[width=\linewidth]{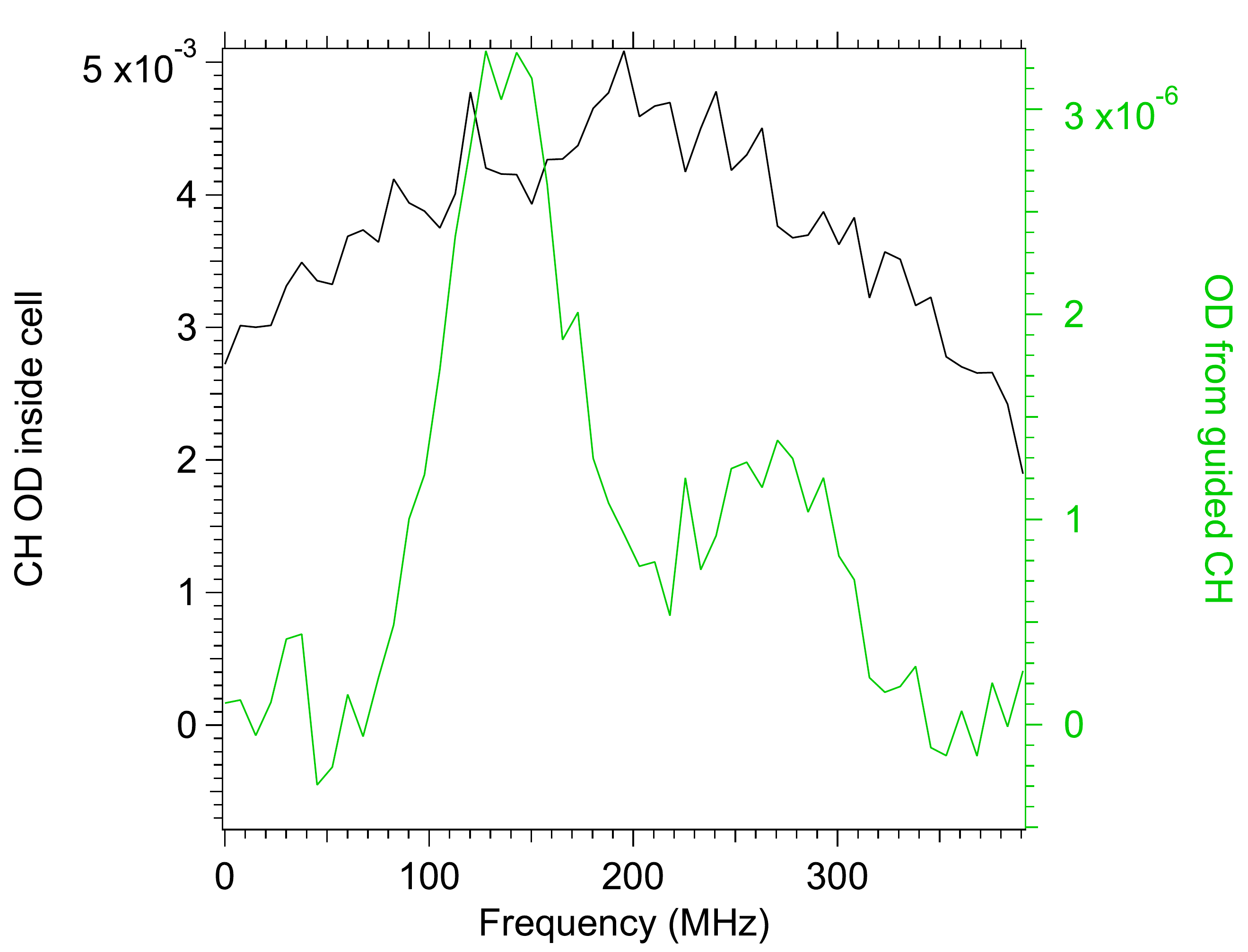}
    					\caption{
    						\label{fig:guided_CH_example} 
    						Spectra of CH inside the cell and at the end of the hexapole with a guiding voltage of 8~kV. Guided OD (green trace) is measured on the right axis, OD inside the cell (black trace) is measured on the left axis.
    						We attribute the two peaks on the guided signal to hyperfine splitting of the excited state of the R$_{22ff}$(1/2) transition, which was previously measured to be 159.6~MHz \cite{CH_microwave_spectroscopy}.
					}     
    			\end{center}
		\end{figure}

In the absence of applying an electrostatic potential to the guide, we observe no fluorescence signal after the guide. Similarly, when probing the R$_{22ee}$(1/2) transition, which probes the high-field-seeking state, we see no observable signal even with voltage applied to the guide. With voltage applied to the hexapole electrodes and exciting the R$_{22ff}$(1/2) transition, we see LIF signals after the guide. Raw signal from the guided CH (plotted along with the in-cell spectrum) is shown in Fig. \ref{fig:guided_CH_example}. We note that there are two visible peaks in our guided spectrum, which we attribute to hyperfine splitting of the R$_{22ff}$(1/2) transition's excited state.
	
\begin{figure}[h!t]
	\begin{center}
		 \includegraphics[width=\linewidth]{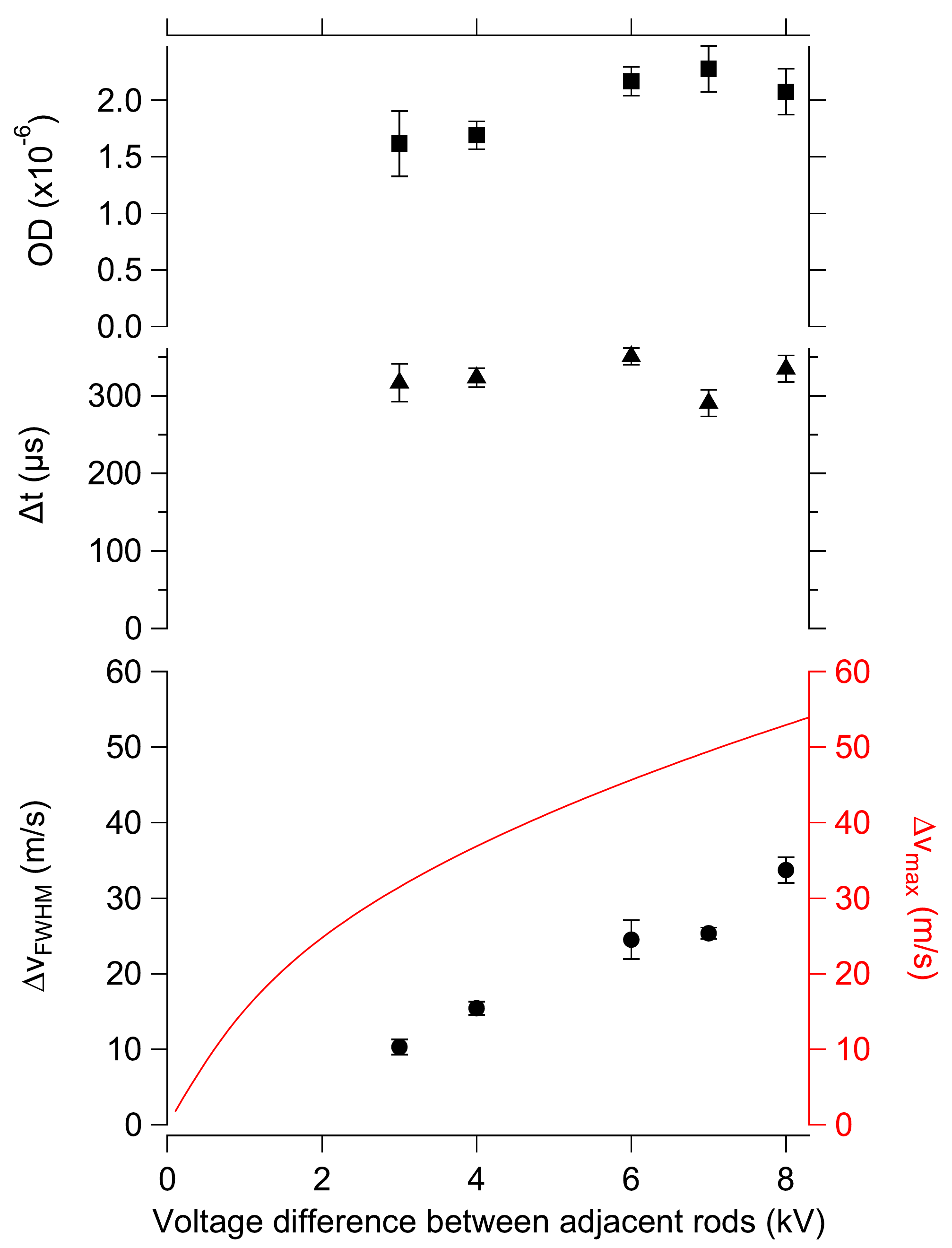}
		\caption{
			\label{fig:guided_CH_measurements}
			 Measurements of the guided CH beam's peak OD, pulse duration, and transverse velocity spread (markers, left axes), plotted as a function of the hexapole voltage. The pulse duration and the transverse velocity spread are both reported as a FWHM.
    		 The red curve (right axis) is the calculated maximum spread in transverse velocities that can be confined by the hexapole field (twice the maximum transverse speed).  
			}     
	\end{center}
\end{figure}

The results of our measurements of the guided beams are shown in Fig. \ref{fig:guided_CH_measurements}. These transverse velocity spreads were calculated from spectra taken 400--600~$\mu$s after ablation, which is the peak of the pulse. The signals are plotted as a function of the voltage difference between adjacent rods of the hexapole guide, with the hexapole rods alternating being positively charged and negatively charged.

We expect that the CH molecules will have a smaller transverse velocity spread after the guide as compared to directly out of the cell for two reasons. First, the transverse confining fields of the guide are weak compared to the expected transverse temperatures of the beam. Second, due to the separation of the guide from the cell (as seen in Fig. \ref{fig:chamber_schematics}) we expect that only molecules with low transverse temperatures will be geometrically accepted. As seen in Fig. \ref{fig:guided_CH_measurements}, this is the case; the transverse velocity spreads correspond to temperatures on the order of a few tenths of a Kelvin. 

Measurements of the peak OD and pulse duration for the guided beam are also shown in Fig. \ref{fig:guided_CH_measurements}.
 %
The guided OD is as high as $\sim$2$\times$10$^{-6}$ (our noise floor in the beam is 10$^{-7}$).
The pulse width is typically measured to be 320$\pm$22~$\mu$s. We calculate the total guided number to be $10^7$ molecules per pulse at our highest guide voltages, and $\sim 10^6$ per pulse for voltages in the 3--4~kV range. We did not see any signal after the guide below 3~kV. 

We note that the number of guided molecules shows a strong dependence on buffer-gas composition, with lower neon fractions producing more guided molecules. The largest numbers were observed when no neon was used at all. This is not what one might expect from measurements of the unguided beam, which shows greater fluxes per solid angle as higher neon flows, as discussed in section \ref{subsec:SmallCellBeam}. 

We also note that only a small fraction of the molecules are actually guided. We speculate that collisions at the entrance of the guide could cause loss of molecules from the beam. Guided CH showed numbers per solid angle per pulse ranging up to $1.5 \times$10$^9$ molecules per steradian.
	
\section{Conclusion} \label{sect:conclusion}

We produced large fluxes of neutral methylidyne radicals at cryogenic temperatures by ablating graphite targets into a mixture of neon buffer gas and hydrogen gas. The lifetime of the molecules in the cell was anomalously short, but by using a small-volume cell with a short extraction time we produced a molecular beam. We successfully guided the neutral methylidyne radical beam using an electrostatic hexapole. Typical molecular beam pulse widths are $\sim 320~\mu$s, measured transverse velocity spreads were 0.1~K (at $\pm$3~kV), and forward velocities of the beam were 250--325~m/s. The total flux of molecules was 1.5$\times 10^9$ molecules per steradian per pulse after the guide.
    
We believe the low total flux of molecules is due to the high temperatures produced by the small-volume cell, and also possibly due to collisional effects at the entrance to the electrostatic guide. We expect improvements in performance could be obtained by returning to a larger volume cell --- which produces colder beams --- and ablating targets with longer lifetimes in the cell. The data presented in Fig. \ref{fig:Target_tau_compar} suggest that with different production methods, this may be possible.
    
\section*{Acknowledgements}	
We thank Heather J. Lewandowski and James Greenberg for the hexapole guide, Stark shift calculations, and experimental assistance.
We thank John M. Doyle and David Patterson for helpful discussions and advice.
This material is based upon work supported by the AFOSR under
Grant No. FA9550-16-1-0117.

\bibliography{CH_writeup_paper_ver}

\begin{thebibliography}{41}%
\makeatletter
\providecommand \@ifxundefined [1]{%
 \@ifx{#1\undefined}
}%
\providecommand \@ifnum [1]{%
 \ifnum #1\expandafter \@firstoftwo
 \else \expandafter \@secondoftwo
 \fi
}%
\providecommand \@ifx [1]{%
 \ifx #1\expandafter \@firstoftwo
 \else \expandafter \@secondoftwo
 \fi
}%
\providecommand \natexlab [1]{#1}%
\providecommand \enquote  [1]{``#1''}%
\providecommand \bibnamefont  [1]{#1}%
\providecommand \bibfnamefont [1]{#1}%
\providecommand \citenamefont [1]{#1}%
\providecommand \href@noop [0]{\@secondoftwo}%
\providecommand \href [0]{\begingroup \@sanitize@url \@href}%
\providecommand \@href[1]{\@@startlink{#1}\@@href}%
\providecommand \@@href[1]{\endgroup#1\@@endlink}%
\providecommand \@sanitize@url [0]{\catcode `\\12\catcode `\$12\catcode
  `\&12\catcode `\#12\catcode `\^12\catcode `\_12\catcode `\%12\relax}%
\providecommand \@@startlink[1]{}%
\providecommand \@@endlink[0]{}%
\providecommand \url  [0]{\begingroup\@sanitize@url \@url }%
\providecommand \@url [1]{\endgroup\@href {#1}{\urlprefix }}%
\providecommand \urlprefix  [0]{URL }%
\providecommand \Eprint [0]{\href }%
\providecommand \doibase [0]{http://dx.doi.org/}%
\providecommand \selectlanguage [0]{\@gobble}%
\providecommand \bibinfo  [0]{\@secondoftwo}%
\providecommand \bibfield  [0]{\@secondoftwo}%
\providecommand \translation [1]{[#1]}%
\providecommand \BibitemOpen [0]{}%
\providecommand \bibitemStop [0]{}%
\providecommand \bibitemNoStop [0]{.\EOS\space}%
\providecommand \EOS [0]{\spacefactor3000\relax}%
\providecommand \BibitemShut  [1]{\csname bibitem#1\endcsname}%
\let\auto@bib@innerbib\@empty
\bibitem [{\citenamefont {Phelps}\ and\ \citenamefont
  {Dalby}(1966)}]{phelps1966experimental}%
  \BibitemOpen
  \bibfield  {author} {\bibinfo {author} {\bibfnamefont {D.}~\bibnamefont
  {Phelps}}\ and\ \bibinfo {author} {\bibfnamefont {F.}~\bibnamefont {Dalby}},\
  }\href@noop {} {\bibfield  {journal} {\bibinfo  {journal} {Physical Review
  Letters}\ }\textbf {\bibinfo {volume} {16}},\ \bibinfo {pages} {3} (\bibinfo
  {year} {1966})}\BibitemShut {NoStop}%
\bibitem [{\citenamefont {McCarthy}\ \emph {et~al.}(2006)\citenamefont
  {McCarthy}, \citenamefont {Mohamed}, \citenamefont {Brown},\ and\
  \citenamefont {Thaddeus}}]{mccarthy2006detection}%
  \BibitemOpen
  \bibfield  {author} {\bibinfo {author} {\bibfnamefont {M.}~\bibnamefont
  {McCarthy}}, \bibinfo {author} {\bibfnamefont {S.}~\bibnamefont {Mohamed}},
  \bibinfo {author} {\bibfnamefont {J.}~\bibnamefont {Brown}}, \ and\ \bibinfo
  {author} {\bibfnamefont {P.}~\bibnamefont {Thaddeus}},\ }\href@noop {}
  {\bibfield  {journal} {\bibinfo  {journal} {Proceedings of the National
  Academy of Sciences}\ }\textbf {\bibinfo {volume} {103}},\ \bibinfo {pages}
  {12263} (\bibinfo {year} {2006})}\BibitemShut {NoStop}%
\bibitem [{\citenamefont {Truppe}\ \emph {et~al.}(2013)\citenamefont {Truppe},
  \citenamefont {Hendricks}, \citenamefont {Tokunaga}, \citenamefont
  {Lewandowski}, \citenamefont {Kozlov}, \citenamefont {Henkel}, \citenamefont
  {Hinds},\ and\ \citenamefont {Tarbutt}}]{truppe2013search}%
  \BibitemOpen
  \bibfield  {author} {\bibinfo {author} {\bibfnamefont {S.}~\bibnamefont
  {Truppe}}, \bibinfo {author} {\bibfnamefont {R.}~\bibnamefont {Hendricks}},
  \bibinfo {author} {\bibfnamefont {S.}~\bibnamefont {Tokunaga}}, \bibinfo
  {author} {\bibfnamefont {H.}~\bibnamefont {Lewandowski}}, \bibinfo {author}
  {\bibfnamefont {M.}~\bibnamefont {Kozlov}}, \bibinfo {author} {\bibfnamefont
  {C.}~\bibnamefont {Henkel}}, \bibinfo {author} {\bibfnamefont
  {E.}~\bibnamefont {Hinds}}, \ and\ \bibinfo {author} {\bibfnamefont
  {M.}~\bibnamefont {Tarbutt}},\ }\href@noop {} {\bibfield  {journal} {\bibinfo
   {journal} {Nature communications}\ }\textbf {\bibinfo {volume} {4}},\
  \bibinfo {pages} {2600} (\bibinfo {year} {2013})}\BibitemShut {NoStop}%
\bibitem [{\citenamefont {Fabrikant}\ \emph {et~al.}(2014)\citenamefont
  {Fabrikant}, \citenamefont {Li}, \citenamefont {Fitch}, \citenamefont
  {Farrow}, \citenamefont {Weinstein},\ and\ \citenamefont
  {Lewandowski}}]{fabrikant2014method}%
  \BibitemOpen
  \bibfield  {author} {\bibinfo {author} {\bibfnamefont {M.}~\bibnamefont
  {Fabrikant}}, \bibinfo {author} {\bibfnamefont {T.}~\bibnamefont {Li}},
  \bibinfo {author} {\bibfnamefont {N.}~\bibnamefont {Fitch}}, \bibinfo
  {author} {\bibfnamefont {N.}~\bibnamefont {Farrow}}, \bibinfo {author}
  {\bibfnamefont {J.~D.}\ \bibnamefont {Weinstein}}, \ and\ \bibinfo {author}
  {\bibfnamefont {H.}~\bibnamefont {Lewandowski}},\ }\href@noop {} {\bibfield
  {journal} {\bibinfo  {journal} {Physical Review A}\ }\textbf {\bibinfo
  {volume} {90}},\ \bibinfo {pages} {033418} (\bibinfo {year}
  {2014})}\BibitemShut {NoStop}%
\bibitem [{\citenamefont {Guang-Bin}\ \emph {et~al.}(2008)\citenamefont
  {Guang-Bin}, \citenamefont {Lian-Zhong},\ and\ \citenamefont
  {Jian-Ping}}]{guang2008new}%
  \BibitemOpen
  \bibfield  {author} {\bibinfo {author} {\bibfnamefont {F.}~\bibnamefont
  {Guang-Bin}}, \bibinfo {author} {\bibfnamefont {D.}~\bibnamefont
  {Lian-Zhong}}, \ and\ \bibinfo {author} {\bibfnamefont {Y.}~\bibnamefont
  {Jian-Ping}},\ }\href@noop {} {\bibfield  {journal} {\bibinfo  {journal}
  {Chinese Physics Letters}\ }\textbf {\bibinfo {volume} {25}},\ \bibinfo
  {pages} {923} (\bibinfo {year} {2008})}\BibitemShut {NoStop}%
\bibitem [{\citenamefont {Railing}\ \emph {et~al.}(2019)\citenamefont
  {Railing}, \citenamefont {Sahw},\ and\ \citenamefont
  {McCarron}}]{DanielMcCarronDAMOP2019}%
  \BibitemOpen
  \bibfield  {author} {\bibinfo {author} {\bibfnamefont {L.}~\bibnamefont
  {Railing}}, \bibinfo {author} {\bibfnamefont {J.}~\bibnamefont {Sahw}}, \
  and\ \bibinfo {author} {\bibfnamefont {D.}~\bibnamefont {McCarron}},\ }in\
  \href@noop {} {\emph {\bibinfo {booktitle} {50th Annual Meeting of the APS
  Division of Atomic, Molecular and Optical Physics}}}\ (\bibinfo {year}
  {2019})\BibitemShut {NoStop}%
\bibitem [{\citenamefont {Bell}\ and\ \citenamefont
  {P.~Softley}(2009)}]{Softley09_UltracoldChemistryReview}%
  \BibitemOpen
  \bibfield  {author} {\bibinfo {author} {\bibfnamefont {M.~T.}\ \bibnamefont
  {Bell}}\ and\ \bibinfo {author} {\bibfnamefont {T.}~\bibnamefont
  {P.~Softley}},\ }\href@noop {} {\bibfield  {journal} {\bibinfo  {journal}
  {Molecular Physics}\ }\textbf {\bibinfo {volume} {107}},\ \bibinfo {pages}
  {99} (\bibinfo {year} {2009})}\BibitemShut {NoStop}%
\bibitem [{\citenamefont {Brownsword}\ \emph {et~al.}(1997)\citenamefont
  {Brownsword}, \citenamefont {Canosa}, \citenamefont {Rowe}, \citenamefont
  {Sims}, \citenamefont {Smith}, \citenamefont {Stewart}, \citenamefont
  {Symonds},\ and\ \citenamefont {Travers}}]{brownsword1997kinetics}%
  \BibitemOpen
  \bibfield  {author} {\bibinfo {author} {\bibfnamefont {R.~A.}\ \bibnamefont
  {Brownsword}}, \bibinfo {author} {\bibfnamefont {A.}~\bibnamefont {Canosa}},
  \bibinfo {author} {\bibfnamefont {B.~R.}\ \bibnamefont {Rowe}}, \bibinfo
  {author} {\bibfnamefont {I.~R.}\ \bibnamefont {Sims}}, \bibinfo {author}
  {\bibfnamefont {I.~W.}\ \bibnamefont {Smith}}, \bibinfo {author}
  {\bibfnamefont {D.~W.}\ \bibnamefont {Stewart}}, \bibinfo {author}
  {\bibfnamefont {A.~C.}\ \bibnamefont {Symonds}}, \ and\ \bibinfo {author}
  {\bibfnamefont {D.}~\bibnamefont {Travers}},\ }\href@noop {} {\bibfield
  {journal} {\bibinfo  {journal} {The Journal of Chemical Physics}\ }\textbf
  {\bibinfo {volume} {106}},\ \bibinfo {pages} {7662} (\bibinfo {year}
  {1997})}\BibitemShut {NoStop}%
\bibitem [{\citenamefont {Maksyutenko}\ \emph {et~al.}(2011)\citenamefont
  {Maksyutenko}, \citenamefont {Zhang}, \citenamefont {Gu},\ and\ \citenamefont
  {Kaiser}}]{maksyutenko2011crossed}%
  \BibitemOpen
  \bibfield  {author} {\bibinfo {author} {\bibfnamefont {P.}~\bibnamefont
  {Maksyutenko}}, \bibinfo {author} {\bibfnamefont {F.}~\bibnamefont {Zhang}},
  \bibinfo {author} {\bibfnamefont {X.}~\bibnamefont {Gu}}, \ and\ \bibinfo
  {author} {\bibfnamefont {R.~I.}\ \bibnamefont {Kaiser}},\ }\href@noop {}
  {\bibfield  {journal} {\bibinfo  {journal} {Physical Chemistry Chemical
  Physics}\ }\textbf {\bibinfo {volume} {13}},\ \bibinfo {pages} {240}
  (\bibinfo {year} {2011})}\BibitemShut {NoStop}%
\bibitem [{\citenamefont {Miller}\ \emph {et~al.}(1990)\citenamefont {Miller},
  \citenamefont {Kee},\ and\ \citenamefont {Westbrook}}]{miller1990chemical}%
  \BibitemOpen
  \bibfield  {author} {\bibinfo {author} {\bibfnamefont {J.~A.}\ \bibnamefont
  {Miller}}, \bibinfo {author} {\bibfnamefont {R.~J.}\ \bibnamefont {Kee}}, \
  and\ \bibinfo {author} {\bibfnamefont {C.~K.}\ \bibnamefont {Westbrook}},\
  }\href@noop {} {\bibfield  {journal} {\bibinfo  {journal} {Annual Review of
  Physical Chemistry}\ }\textbf {\bibinfo {volume} {41}},\ \bibinfo {pages}
  {345} (\bibinfo {year} {1990})}\BibitemShut {NoStop}%
\bibitem [{\citenamefont {Canosa}\ \emph {et~al.}(1997)\citenamefont {Canosa},
  \citenamefont {Sims}, \citenamefont {Travers}, \citenamefont {Smith},\ and\
  \citenamefont {Rowe}}]{canosa1997reactions}%
  \BibitemOpen
  \bibfield  {author} {\bibinfo {author} {\bibfnamefont {A.}~\bibnamefont
  {Canosa}}, \bibinfo {author} {\bibfnamefont {I.~R.}\ \bibnamefont {Sims}},
  \bibinfo {author} {\bibfnamefont {D.}~\bibnamefont {Travers}}, \bibinfo
  {author} {\bibfnamefont {I.~W.}\ \bibnamefont {Smith}}, \ and\ \bibinfo
  {author} {\bibfnamefont {B.}~\bibnamefont {Rowe}},\ }\href@noop {} {\bibfield
   {journal} {\bibinfo  {journal} {Astronomy and Astrophysics}\ }\textbf
  {\bibinfo {volume} {323}},\ \bibinfo {pages} {644} (\bibinfo {year}
  {1997})}\BibitemShut {NoStop}%
\bibitem [{\citenamefont {Prasad}\ and\ \citenamefont
  {Huntress~Jr}(1980)}]{prasad1980model}%
  \BibitemOpen
  \bibfield  {author} {\bibinfo {author} {\bibfnamefont {S.}~\bibnamefont
  {Prasad}}\ and\ \bibinfo {author} {\bibfnamefont {W.}~\bibnamefont
  {Huntress~Jr}},\ }\href@noop {} {\bibfield  {journal} {\bibinfo  {journal}
  {The Astrophysical Journal Supplement Series}\ }\textbf {\bibinfo {volume}
  {43}},\ \bibinfo {pages} {1} (\bibinfo {year} {1980})}\BibitemShut {NoStop}%
\bibitem [{\citenamefont {Millar}\ and\ \citenamefont
  {Williams}(1991)}]{millar1991interstellar}%
  \BibitemOpen
  \bibfield  {author} {\bibinfo {author} {\bibfnamefont {T.}~\bibnamefont
  {Millar}}\ and\ \bibinfo {author} {\bibfnamefont {D.}~\bibnamefont
  {Williams}},\ }\href@noop {} {\bibfield  {journal} {\bibinfo  {journal}
  {Science Progress (1933-)}\ ,\ \bibinfo {pages} {279}} (\bibinfo {year}
  {1991})}\BibitemShut {NoStop}%
\bibitem [{\citenamefont {Henson}\ \emph {et~al.}(2012)\citenamefont {Henson},
  \citenamefont {Gersten}, \citenamefont {Shagam}, \citenamefont {Narevicius},\
  and\ \citenamefont {Narevicius}}]{Henson234}%
  \BibitemOpen
  \bibfield  {author} {\bibinfo {author} {\bibfnamefont {A.~B.}\ \bibnamefont
  {Henson}}, \bibinfo {author} {\bibfnamefont {S.}~\bibnamefont {Gersten}},
  \bibinfo {author} {\bibfnamefont {Y.}~\bibnamefont {Shagam}}, \bibinfo
  {author} {\bibfnamefont {J.}~\bibnamefont {Narevicius}}, \ and\ \bibinfo
  {author} {\bibfnamefont {E.}~\bibnamefont {Narevicius}},\ }\href {\doibase
  10.1126/science.1229141} {\bibfield  {journal} {\bibinfo  {journal}
  {Science}\ }\textbf {\bibinfo {volume} {338}},\ \bibinfo {pages} {234}
  (\bibinfo {year} {2012})}\BibitemShut {NoStop}%
\bibitem [{\citenamefont {Greenberg}\ \emph {et~al.}(2018)\citenamefont
  {Greenberg}, \citenamefont {Schmid}, \citenamefont {Miller}, \citenamefont
  {Stanton},\ and\ \citenamefont {Lewandowski}}]{greenberg2018quantum}%
  \BibitemOpen
  \bibfield  {author} {\bibinfo {author} {\bibfnamefont {J.}~\bibnamefont
  {Greenberg}}, \bibinfo {author} {\bibfnamefont {P.~C.}\ \bibnamefont
  {Schmid}}, \bibinfo {author} {\bibfnamefont {M.}~\bibnamefont {Miller}},
  \bibinfo {author} {\bibfnamefont {J.~F.}\ \bibnamefont {Stanton}}, \ and\
  \bibinfo {author} {\bibfnamefont {H.}~\bibnamefont {Lewandowski}},\
  }\href@noop {} {\bibfield  {journal} {\bibinfo  {journal} {Physical Review
  A}\ }\textbf {\bibinfo {volume} {98}},\ \bibinfo {pages} {032702} (\bibinfo
  {year} {2018})}\BibitemShut {NoStop}%
\bibitem [{\citenamefont {Chen}\ \emph {et~al.}(2019)\citenamefont {Chen},
  \citenamefont {Xie}, \citenamefont {Yang}, \citenamefont {Li}, \citenamefont
  {Suits}, \citenamefont {Hudson}, \citenamefont {Campbell},\ and\
  \citenamefont {Guo}}]{chen2019isotope}%
  \BibitemOpen
  \bibfield  {author} {\bibinfo {author} {\bibfnamefont {G.~K.}\ \bibnamefont
  {Chen}}, \bibinfo {author} {\bibfnamefont {C.}~\bibnamefont {Xie}}, \bibinfo
  {author} {\bibfnamefont {T.}~\bibnamefont {Yang}}, \bibinfo {author}
  {\bibfnamefont {A.}~\bibnamefont {Li}}, \bibinfo {author} {\bibfnamefont
  {A.~G.}\ \bibnamefont {Suits}}, \bibinfo {author} {\bibfnamefont {E.~R.}\
  \bibnamefont {Hudson}}, \bibinfo {author} {\bibfnamefont {W.~C.}\
  \bibnamefont {Campbell}}, \ and\ \bibinfo {author} {\bibfnamefont
  {H.}~\bibnamefont {Guo}},\ }\href@noop {} {\bibfield  {journal} {\bibinfo
  {journal} {Physical Chemistry Chemical Physics}\ } (\bibinfo {year}
  {2019})}\BibitemShut {NoStop}%
\bibitem [{\citenamefont {Maxwell}\ \emph {et~al.}(2005)\citenamefont
  {Maxwell}, \citenamefont {Brahms}, \citenamefont {Glenn}, \citenamefont
  {Helton}, \citenamefont {Nguyen}, \citenamefont {Patterson}, \citenamefont
  {Petricka}, \citenamefont {DeMille}, \citenamefont {Doyle} \emph
  {et~al.}}]{maxwell2005high}%
  \BibitemOpen
  \bibfield  {author} {\bibinfo {author} {\bibfnamefont {S.~E.}\ \bibnamefont
  {Maxwell}}, \bibinfo {author} {\bibfnamefont {N.}~\bibnamefont {Brahms}},
  \bibinfo {author} {\bibfnamefont {D.}~\bibnamefont {Glenn}}, \bibinfo
  {author} {\bibfnamefont {J.}~\bibnamefont {Helton}}, \bibinfo {author}
  {\bibfnamefont {S.}~\bibnamefont {Nguyen}}, \bibinfo {author} {\bibfnamefont
  {D.}~\bibnamefont {Patterson}}, \bibinfo {author} {\bibfnamefont
  {J.}~\bibnamefont {Petricka}}, \bibinfo {author} {\bibfnamefont
  {D.}~\bibnamefont {DeMille}}, \bibinfo {author} {\bibfnamefont
  {J.}~\bibnamefont {Doyle}},  \emph {et~al.},\ }\href@noop {} {\bibfield
  {journal} {\bibinfo  {journal} {Physical Review Letters}\ }\textbf {\bibinfo
  {volume} {95}},\ \bibinfo {pages} {173201} (\bibinfo {year}
  {2005})}\BibitemShut {NoStop}%
\bibitem [{\citenamefont {Bethlem}\ \emph {et~al.}(1999)\citenamefont
  {Bethlem}, \citenamefont {Berden},\ and\ \citenamefont
  {Meijer}}]{PhysRevLett.83.1558}%
  \BibitemOpen
  \bibfield  {author} {\bibinfo {author} {\bibfnamefont {H.~L.}\ \bibnamefont
  {Bethlem}}, \bibinfo {author} {\bibfnamefont {G.}~\bibnamefont {Berden}}, \
  and\ \bibinfo {author} {\bibfnamefont {G.}~\bibnamefont {Meijer}},\ }\href
  {\doibase 10.1103/PhysRevLett.83.1558} {\bibfield  {journal} {\bibinfo
  {journal} {Physical Review Letters}\ }\textbf {\bibinfo {volume} {83}},\
  \bibinfo {pages} {1558} (\bibinfo {year} {1999})}\BibitemShut {NoStop}%
\bibitem [{\citenamefont {Barry}\ \emph {et~al.}(2014)\citenamefont {Barry},
  \citenamefont {McCarron}, \citenamefont {Norrgard}, \citenamefont
  {Steinecker},\ and\ \citenamefont {DeMille}}]{barry2014magneto}%
  \BibitemOpen
  \bibfield  {author} {\bibinfo {author} {\bibfnamefont {J.}~\bibnamefont
  {Barry}}, \bibinfo {author} {\bibfnamefont {D.}~\bibnamefont {McCarron}},
  \bibinfo {author} {\bibfnamefont {E.}~\bibnamefont {Norrgard}}, \bibinfo
  {author} {\bibfnamefont {M.}~\bibnamefont {Steinecker}}, \ and\ \bibinfo
  {author} {\bibfnamefont {D.}~\bibnamefont {DeMille}},\ }\href@noop {}
  {\bibfield  {journal} {\bibinfo  {journal} {Nature}\ }\textbf {\bibinfo
  {volume} {512}},\ \bibinfo {pages} {286} (\bibinfo {year}
  {2014})}\BibitemShut {NoStop}%
\bibitem [{\citenamefont {Andreev}\ \emph {et~al.}(2018)\citenamefont
  {Andreev}, \citenamefont {Ang}, \citenamefont {DeMille}, \citenamefont
  {Doyle}, \citenamefont {Gabrielse}, \citenamefont {Haefner}, \citenamefont
  {Hutzler}, \citenamefont {Lasner}, \citenamefont {Meisenhelder},
  \citenamefont {O'Leary} \emph {et~al.}}]{andreev2018improved}%
  \BibitemOpen
  \bibfield  {author} {\bibinfo {author} {\bibfnamefont {V.}~\bibnamefont
  {Andreev}}, \bibinfo {author} {\bibfnamefont {D.}~\bibnamefont {Ang}},
  \bibinfo {author} {\bibfnamefont {D.}~\bibnamefont {DeMille}}, \bibinfo
  {author} {\bibfnamefont {J.}~\bibnamefont {Doyle}}, \bibinfo {author}
  {\bibfnamefont {G.}~\bibnamefont {Gabrielse}}, \bibinfo {author}
  {\bibfnamefont {J.}~\bibnamefont {Haefner}}, \bibinfo {author} {\bibfnamefont
  {N.}~\bibnamefont {Hutzler}}, \bibinfo {author} {\bibfnamefont
  {Z.}~\bibnamefont {Lasner}}, \bibinfo {author} {\bibfnamefont
  {C.}~\bibnamefont {Meisenhelder}}, \bibinfo {author} {\bibfnamefont
  {B.}~\bibnamefont {O'Leary}},  \emph {et~al.},\ }\href@noop {} {\bibfield
  {journal} {\bibinfo  {journal} {Nature}\ }\textbf {\bibinfo {volume} {562}},\
  \bibinfo {pages} {355} (\bibinfo {year} {2018})}\BibitemShut {NoStop}%
\bibitem [{\citenamefont {van~de Meerakker}\ \emph {et~al.}(2012)\citenamefont
  {van~de Meerakker}, \citenamefont {Bethlem}, \citenamefont {Vanhaecke},\ and\
  \citenamefont {Meijer}}]{van2012manipulation}%
  \BibitemOpen
  \bibfield  {author} {\bibinfo {author} {\bibfnamefont {S.~Y.}\ \bibnamefont
  {van~de Meerakker}}, \bibinfo {author} {\bibfnamefont {H.~L.}\ \bibnamefont
  {Bethlem}}, \bibinfo {author} {\bibfnamefont {N.}~\bibnamefont {Vanhaecke}},
  \ and\ \bibinfo {author} {\bibfnamefont {G.}~\bibnamefont {Meijer}},\
  }\href@noop {} {\bibfield  {journal} {\bibinfo  {journal} {Chemical Reviews}\
  }\textbf {\bibinfo {volume} {112}},\ \bibinfo {pages} {4828} (\bibinfo {year}
  {2012})}\BibitemShut {NoStop}%
\bibitem [{\citenamefont {Patterson}\ and\ \citenamefont
  {Doyle}(2007)}]{patterson2007bright}%
  \BibitemOpen
  \bibfield  {author} {\bibinfo {author} {\bibfnamefont {D.}~\bibnamefont
  {Patterson}}\ and\ \bibinfo {author} {\bibfnamefont {J.~M.}\ \bibnamefont
  {Doyle}},\ }\href@noop {} {\bibfield  {journal} {\bibinfo  {journal} {The
  Journal of Chemical Physics}\ }\textbf {\bibinfo {volume} {126}},\ \bibinfo
  {pages} {154307} (\bibinfo {year} {2007})}\BibitemShut {NoStop}%
\bibitem [{\citenamefont {van Buuren}\ \emph {et~al.}(2009)\citenamefont {van
  Buuren}, \citenamefont {Sommer}, \citenamefont {Motsch}, \citenamefont
  {Pohle}, \citenamefont {Schenk}, \citenamefont {Bayerl}, \citenamefont
  {Pinkse},\ and\ \citenamefont {Rempe}}]{van2009electrostatic}%
  \BibitemOpen
  \bibfield  {author} {\bibinfo {author} {\bibfnamefont {L.~D.}\ \bibnamefont
  {van Buuren}}, \bibinfo {author} {\bibfnamefont {C.}~\bibnamefont {Sommer}},
  \bibinfo {author} {\bibfnamefont {M.}~\bibnamefont {Motsch}}, \bibinfo
  {author} {\bibfnamefont {S.}~\bibnamefont {Pohle}}, \bibinfo {author}
  {\bibfnamefont {M.}~\bibnamefont {Schenk}}, \bibinfo {author} {\bibfnamefont
  {J.}~\bibnamefont {Bayerl}}, \bibinfo {author} {\bibfnamefont {P.~W.}\
  \bibnamefont {Pinkse}}, \ and\ \bibinfo {author} {\bibfnamefont
  {G.}~\bibnamefont {Rempe}},\ }\href@noop {} {\bibfield  {journal} {\bibinfo
  {journal} {Physical Review Letters}\ }\textbf {\bibinfo {volume} {102}},\
  \bibinfo {pages} {033001} (\bibinfo {year} {2009})}\BibitemShut {NoStop}%
\bibitem [{\citenamefont {Patterson}\ \emph {et~al.}(2009)\citenamefont
  {Patterson}, \citenamefont {Rasmussen},\ and\ \citenamefont
  {Doyle}}]{patterson2009intense}%
  \BibitemOpen
  \bibfield  {author} {\bibinfo {author} {\bibfnamefont {D.}~\bibnamefont
  {Patterson}}, \bibinfo {author} {\bibfnamefont {J.}~\bibnamefont
  {Rasmussen}}, \ and\ \bibinfo {author} {\bibfnamefont {J.~M.}\ \bibnamefont
  {Doyle}},\ }\href@noop {} {\bibfield  {journal} {\bibinfo  {journal} {New
  Journal of Physics}\ }\textbf {\bibinfo {volume} {11}},\ \bibinfo {pages}
  {055018} (\bibinfo {year} {2009})}\BibitemShut {NoStop}%
\bibitem [{\citenamefont {Staanum}\ \emph {et~al.}(2008)\citenamefont
  {Staanum}, \citenamefont {H{\o}jbjerre}, \citenamefont {Wester},\ and\
  \citenamefont {Drewsen}}]{staanum2008probing}%
  \BibitemOpen
  \bibfield  {author} {\bibinfo {author} {\bibfnamefont {P.~F.}\ \bibnamefont
  {Staanum}}, \bibinfo {author} {\bibfnamefont {K.}~\bibnamefont
  {H{\o}jbjerre}}, \bibinfo {author} {\bibfnamefont {R.}~\bibnamefont
  {Wester}}, \ and\ \bibinfo {author} {\bibfnamefont {M.}~\bibnamefont
  {Drewsen}},\ }\href@noop {} {\bibfield  {journal} {\bibinfo  {journal}
  {Physical Review Letters}\ }\textbf {\bibinfo {volume} {100}},\ \bibinfo
  {pages} {243003} (\bibinfo {year} {2008})}\BibitemShut {NoStop}%
\bibitem [{\citenamefont {Willitsch}\ \emph {et~al.}(2008)\citenamefont
  {Willitsch}, \citenamefont {Bell}, \citenamefont {Gingell}, \citenamefont
  {Procter},\ and\ \citenamefont {Softley}}]{willitsch2008cold}%
  \BibitemOpen
  \bibfield  {author} {\bibinfo {author} {\bibfnamefont {S.}~\bibnamefont
  {Willitsch}}, \bibinfo {author} {\bibfnamefont {M.~T.}\ \bibnamefont {Bell}},
  \bibinfo {author} {\bibfnamefont {A.~D.}\ \bibnamefont {Gingell}}, \bibinfo
  {author} {\bibfnamefont {S.~R.}\ \bibnamefont {Procter}}, \ and\ \bibinfo
  {author} {\bibfnamefont {T.~P.}\ \bibnamefont {Softley}},\ }\href@noop {}
  {\bibfield  {journal} {\bibinfo  {journal} {Physical Review Letters}\
  }\textbf {\bibinfo {volume} {100}},\ \bibinfo {pages} {043203} (\bibinfo
  {year} {2008})}\BibitemShut {NoStop}%
\bibitem [{\citenamefont {Weinstein}\ \emph
  {et~al.}(1998{\natexlab{a}})\citenamefont {Weinstein}, \citenamefont
  {Decarvalho}, \citenamefont {Amar}, \citenamefont {Boca}, \citenamefont
  {Odom}, \citenamefont {Friedrich},\ and\ \citenamefont
  {Doyle}}]{weinstein1998spectroscopy}%
  \BibitemOpen
  \bibfield  {author} {\bibinfo {author} {\bibfnamefont {J.~D.}\ \bibnamefont
  {Weinstein}}, \bibinfo {author} {\bibfnamefont {R.}~\bibnamefont
  {Decarvalho}}, \bibinfo {author} {\bibfnamefont {K.}~\bibnamefont {Amar}},
  \bibinfo {author} {\bibfnamefont {A.}~\bibnamefont {Boca}}, \bibinfo {author}
  {\bibfnamefont {B.~C.}\ \bibnamefont {Odom}}, \bibinfo {author}
  {\bibfnamefont {B.}~\bibnamefont {Friedrich}}, \ and\ \bibinfo {author}
  {\bibfnamefont {J.~M.}\ \bibnamefont {Doyle}},\ }\href@noop {} {\bibfield
  {journal} {\bibinfo  {journal} {The Journal of Chemical Physics}\ }\textbf
  {\bibinfo {volume} {109}},\ \bibinfo {pages} {2656} (\bibinfo {year}
  {1998}{\natexlab{a}})}\BibitemShut {NoStop}%
\bibitem [{\citenamefont {Weinstein}\ \emph
  {et~al.}(1998{\natexlab{b}})\citenamefont {Weinstein}, \citenamefont
  {deCarvalho}, \citenamefont {Guillet}, \citenamefont {Friedrich},\ and\
  \citenamefont {Doyle}}]{CaHnature}%
  \BibitemOpen
  \bibfield  {author} {\bibinfo {author} {\bibfnamefont {J.~D.}\ \bibnamefont
  {Weinstein}}, \bibinfo {author} {\bibfnamefont {R.}~\bibnamefont
  {deCarvalho}}, \bibinfo {author} {\bibfnamefont {T.}~\bibnamefont {Guillet}},
  \bibinfo {author} {\bibfnamefont {B.}~\bibnamefont {Friedrich}}, \ and\
  \bibinfo {author} {\bibfnamefont {J.~M.}\ \bibnamefont {Doyle}},\ }\href@noop
  {} {\bibfield  {journal} {\bibinfo  {journal} {Nature}\ }\textbf {\bibinfo
  {volume} {395}},\ \bibinfo {pages} {148} (\bibinfo {year}
  {1998}{\natexlab{b}})}\BibitemShut {NoStop}%
\bibitem [{\citenamefont {{Truppe}}\ \emph {et~al.}(2018)\citenamefont
  {{Truppe}}, \citenamefont {{Hambach}}, \citenamefont {{Skoff}}, \citenamefont
  {{Bulleid}}, \citenamefont {{Bumby}}, \citenamefont {{Hendricks}},
  \citenamefont {{Hinds}}, \citenamefont {{Sauer}},\ and\ \citenamefont
  {{Tarbutt}}}]{ImperialCollege2017}%
  \BibitemOpen
  \bibfield  {author} {\bibinfo {author} {\bibfnamefont {S.}~\bibnamefont
  {{Truppe}}}, \bibinfo {author} {\bibfnamefont {M.}~\bibnamefont {{Hambach}}},
  \bibinfo {author} {\bibfnamefont {S.~M.}\ \bibnamefont {{Skoff}}}, \bibinfo
  {author} {\bibfnamefont {N.~E.}\ \bibnamefont {{Bulleid}}}, \bibinfo {author}
  {\bibfnamefont {J.~S.}\ \bibnamefont {{Bumby}}}, \bibinfo {author}
  {\bibfnamefont {R.~J.}\ \bibnamefont {{Hendricks}}}, \bibinfo {author}
  {\bibfnamefont {E.~A.}\ \bibnamefont {{Hinds}}}, \bibinfo {author}
  {\bibfnamefont {B.~E.}\ \bibnamefont {{Sauer}}}, \ and\ \bibinfo {author}
  {\bibfnamefont {M.~R.}\ \bibnamefont {{Tarbutt}}},\ }\href@noop {} {\bibfield
   {journal} {\bibinfo  {journal} {Journal of Modern Optics}\ }\textbf
  {\bibinfo {volume} {65}},\ \bibinfo {pages} {648} (\bibinfo {year}
  {2018})}\BibitemShut {NoStop}%
\bibitem [{\citenamefont {Zachwieja}(1995)}]{CH_wavenum__ZACHWIEJA1995285}%
  \BibitemOpen
  \bibfield  {author} {\bibinfo {author} {\bibfnamefont {M.}~\bibnamefont
  {Zachwieja}},\ }\href {\doibase https://doi.org/10.1006/jmsp.1995.1072}
  {\bibfield  {journal} {\bibinfo  {journal} {Journal of Molecular
  Spectroscopy}\ }\textbf {\bibinfo {volume} {170}},\ \bibinfo {pages} {285 }
  (\bibinfo {year} {1995})}\BibitemShut {NoStop}%
\bibitem [{\citenamefont {Truppe}\ \emph {et~al.}(2014)\citenamefont {Truppe},
  \citenamefont {Hendricks}, \citenamefont {Tokunaga}, \citenamefont {Hinds},\
  and\ \citenamefont {Tarbutt}}]{CH_microwave_spectroscopy}%
  \BibitemOpen
  \bibfield  {author} {\bibinfo {author} {\bibfnamefont {S.}~\bibnamefont
  {Truppe}}, \bibinfo {author} {\bibfnamefont {R.}~\bibnamefont {Hendricks}},
  \bibinfo {author} {\bibfnamefont {S.}~\bibnamefont {Tokunaga}}, \bibinfo
  {author} {\bibfnamefont {E.}~\bibnamefont {Hinds}}, \ and\ \bibinfo {author}
  {\bibfnamefont {M.}~\bibnamefont {Tarbutt}},\ }\href@noop {} {\bibfield
  {journal} {\bibinfo  {journal} {Journal of Molecular Spectroscopy}\ }\textbf
  {\bibinfo {volume} {300}} (\bibinfo {year} {2014})}\BibitemShut {NoStop}%
\bibitem [{\citenamefont {Luque}\ and\ \citenamefont
  {Crosley}(1996)}]{CH_transition_probabilities}%
  \BibitemOpen
  \bibfield  {author} {\bibinfo {author} {\bibfnamefont {J.}~\bibnamefont
  {Luque}}\ and\ \bibinfo {author} {\bibfnamefont {D.~R.}\ \bibnamefont
  {Crosley}},\ }\href@noop {} {\bibfield  {journal} {\bibinfo  {journal} {The
  Journal of Chemical Physics}\ }\textbf {\bibinfo {volume} {104}},\ \bibinfo
  {pages} {2146} (\bibinfo {year} {1996})}\BibitemShut {NoStop}%
\bibitem [{\citenamefont {Budker}\ \emph {et~al.}(2004)\citenamefont {Budker},
  \citenamefont {Kimball},\ and\ \citenamefont {DeMille}}]{BudkerBook}%
  \BibitemOpen
  \bibfield  {author} {\bibinfo {author} {\bibfnamefont {D.}~\bibnamefont
  {Budker}}, \bibinfo {author} {\bibfnamefont {D.~F.}\ \bibnamefont {Kimball}},
  \ and\ \bibinfo {author} {\bibfnamefont {D.}~\bibnamefont {DeMille}},\
  }\href@noop {} {\emph {\bibinfo {title} {Atomic Physics}}}\ (\bibinfo
  {publisher} {Oxford University Press},\ \bibinfo {year} {2004})\BibitemShut
  {NoStop}%
\bibitem [{\citenamefont {Lancaster}(2018)}]{Lancaster2018Thesis}%
  \BibitemOpen
  \bibfield  {author} {\bibinfo {author} {\bibfnamefont {D.}~\bibnamefont
  {Lancaster}},\ }\emph {\bibinfo {title} {``A Source of Cold Molecules:
  Cryostat Design, Construction, and Finding a CH Source"}},\ \href
  {http://www.physics.unr.edu/xap/publications.html} {\bibinfo {type} {{B.S.
  Thesis}}},\ \bibinfo  {school} {University of Nevada, Reno} (\bibinfo {year}
  {2018})\BibitemShut {NoStop}%
\bibitem [{\citenamefont {Xiao}\ \emph {et~al.}(2019)\citenamefont {Xiao},
  \citenamefont {Lancaster}, \citenamefont {Allen}, \citenamefont {Taylor},
  \citenamefont {Lancaster}, \citenamefont {Shaw}, \citenamefont {Hutzler},\
  and\ \citenamefont {Weinstein}}]{Nozzles2018}%
  \BibitemOpen
  \bibfield  {author} {\bibinfo {author} {\bibfnamefont {D.}~\bibnamefont
  {Xiao}}, \bibinfo {author} {\bibfnamefont {D.~M.}\ \bibnamefont {Lancaster}},
  \bibinfo {author} {\bibfnamefont {C.~H.}\ \bibnamefont {Allen}}, \bibinfo
  {author} {\bibfnamefont {M.~J.}\ \bibnamefont {Taylor}}, \bibinfo {author}
  {\bibfnamefont {T.~A.}\ \bibnamefont {Lancaster}}, \bibinfo {author}
  {\bibfnamefont {G.}~\bibnamefont {Shaw}}, \bibinfo {author} {\bibfnamefont
  {N.~R.}\ \bibnamefont {Hutzler}}, \ and\ \bibinfo {author} {\bibfnamefont
  {J.~D.}\ \bibnamefont {Weinstein}},\ }\href {\doibase
  10.1103/PhysRevA.99.013603} {\bibfield  {journal} {\bibinfo  {journal} {Phys.
  Rev. A}\ }\textbf {\bibinfo {volume} {99}},\ \bibinfo {pages} {013603}
  (\bibinfo {year} {2019})}\BibitemShut {NoStop}%
\bibitem [{\citenamefont {Taylor}(2018)}]{MackTaylorThesis2018}%
  \BibitemOpen
  \bibfield  {author} {\bibinfo {author} {\bibfnamefont {M.}~\bibnamefont
  {Taylor}},\ }\emph {\bibinfo {title} {Characterizing a Cold Molecular Beam
  Source Using de Laval Nozzles}},\ \href
  {http://www.physics.unr.edu/xap/publications.html} {\bibinfo {type} {{B.S.
  Thesis}}},\ \bibinfo  {school} {University of Nevada, Reno} (\bibinfo {year}
  {2018})\BibitemShut {NoStop}%
\bibitem [{\citenamefont {Linstrom}\ and\ \citenamefont
  {Mallard}(2019)}]{NISTChemWebBook}%
  \BibitemOpen
  \bibinfo {editor} {\bibfnamefont {P.~J.}\ \bibnamefont {Linstrom}}\ and\
  \bibinfo {editor} {\bibfnamefont {W.~G.}\ \bibnamefont {Mallard}},\ eds.,\
  \href@noop {} {\emph {\bibinfo {title} {NIST Chemistry WebBook, NIST Standard
  Reference database Number 69}}}\ (\bibinfo  {publisher} {National Institute
  of Standards and Technology},\ \bibinfo {address} {Gaithersburg, MD 20899},\
  \bibinfo {year} {2019})\ \bibinfo {note}
  {http://webbook.nist.gov/}\BibitemShut {NoStop}%
\bibitem [{\citenamefont {Quiros}\ \emph {et~al.}(2017)\citenamefont {Quiros},
  \citenamefont {Tariq}, \citenamefont {Tscherbul}, \citenamefont {K{\l}os},\
  and\ \citenamefont {Weinstein}}]{quiros2017cold}%
  \BibitemOpen
  \bibfield  {author} {\bibinfo {author} {\bibfnamefont {N.}~\bibnamefont
  {Quiros}}, \bibinfo {author} {\bibfnamefont {N.}~\bibnamefont {Tariq}},
  \bibinfo {author} {\bibfnamefont {T.~V.}\ \bibnamefont {Tscherbul}}, \bibinfo
  {author} {\bibfnamefont {J.}~\bibnamefont {K{\l}os}}, \ and\ \bibinfo
  {author} {\bibfnamefont {J.~D.}\ \bibnamefont {Weinstein}},\ }\href@noop {}
  {\bibfield  {journal} {\bibinfo  {journal} {Physical Review Letters}\
  }\textbf {\bibinfo {volume} {118}},\ \bibinfo {pages} {213401} (\bibinfo
  {year} {2017})}\BibitemShut {NoStop}%
\bibitem [{\citenamefont {Lu}\ \emph {et~al.}(2009)\citenamefont {Lu},
  \citenamefont {Singh},\ and\ \citenamefont {Weinstein}}]{lu2009inelastic}%
  \BibitemOpen
  \bibfield  {author} {\bibinfo {author} {\bibfnamefont {M.-J.}\ \bibnamefont
  {Lu}}, \bibinfo {author} {\bibfnamefont {V.}~\bibnamefont {Singh}}, \ and\
  \bibinfo {author} {\bibfnamefont {J.~D.}\ \bibnamefont {Weinstein}},\
  }\href@noop {} {\bibfield  {journal} {\bibinfo  {journal} {Physical Review
  A}\ }\textbf {\bibinfo {volume} {79}},\ \bibinfo {pages} {050702} (\bibinfo
  {year} {2009})}\BibitemShut {NoStop}%
\bibitem [{\citenamefont {Sansonetti}\ \emph {et~al.}(2005)\citenamefont
  {Sansonetti}, \citenamefont {Martin},\ and\ \citenamefont
  {Young}}]{NISTAtomicData}%
  \BibitemOpen
  \bibfield  {author} {\bibinfo {author} {\bibfnamefont {J.~E.}\ \bibnamefont
  {Sansonetti}}, \bibinfo {author} {\bibfnamefont {W.~C.}\ \bibnamefont
  {Martin}}, \ and\ \bibinfo {author} {\bibfnamefont {S.~L.}\ \bibnamefont
  {Young}},\ }\href@noop {} {\emph {\bibinfo {title} {Handbook of Basic Atomic
  Spectroscopic Data (version 1.1.2)}}}\ (\bibinfo  {publisher} {National
  Institute of Standards and Technology, Gaithersburg, MD},\ \bibinfo {year}
  {2005})\ \bibinfo {note} {http://physics.nist.gov/Handbook}\BibitemShut
  {NoStop}%
\bibitem [{\citenamefont {Hutzler}\ \emph {et~al.}(2012)\citenamefont
  {Hutzler}, \citenamefont {Lu},\ and\ \citenamefont
  {Doyle}}]{hutzler2012buffer}%
  \BibitemOpen
  \bibfield  {author} {\bibinfo {author} {\bibfnamefont {N.~R.}\ \bibnamefont
  {Hutzler}}, \bibinfo {author} {\bibfnamefont {H.-I.}\ \bibnamefont {Lu}}, \
  and\ \bibinfo {author} {\bibfnamefont {J.~M.}\ \bibnamefont {Doyle}},\
  }\href@noop {} {\bibfield  {journal} {\bibinfo  {journal} {Chemical reviews}\
  }\textbf {\bibinfo {volume} {112}},\ \bibinfo {pages} {4803} (\bibinfo {year}
  {2012})}\BibitemShut {NoStop}%
\end{thebibliography}%

\appendix

\section{Titanium measurements} \label{sect:TiResults}
	
    \subsection{Standard cell} \label{subsec:StandardCellTiResults}
Initial tests of the cryogenic buffer-gas beam were conducted with titanium atoms produced by ablation of titanium metal. Titanium was chosen for its ease of production by ablation and its strong and accessible optical transition \cite{lu2009inelastic}. Atoms are created through laser ablation with a frequency-doubled Nd:YAG laser operating at 10~Hz with energies of 10--80~mJ per pulse. The atoms are detected with a 400~nm tuneable diode laser driving the $3d^24s^2\, a\, ^3\mathrm{F}_2 \rightarrow 3d^24s4p\, y\, ^3 \mathrm{F}^{\circ}_2$ transition at 25107.4~cm$^{-1}$ \cite{NISTAtomicData}, with typical powers of 10--20~$\mu$W and beam waist of a few~mm. Production of Ti is high, yielding  3$\times$10$^{13}$ atoms in the ground state. We note that at early times the OD is often saturated, preventing it from being measuring directly. In such cases, peak values are extrapolated from the unsaturated data assuming an exponential decay in signal, which is observed to be a good approximation when peak values can be directly observed, as seen in Fig. \ref{fig:Target_tau_compar}. 
		
We fit the measured OD to an exponential decay to extract a lifetime. At low neon buffer-gas flow rates ($\gtrsim 4$~sccm and $\lesssim 40$~sccm) we observe the titanium lifetime to increase linearly with increasing buffer-gas flow rate. This is as expected, as diffusion to the cell walls is the dominant loss process at low buffer-gas densities \cite{hutzler2012buffer}. At higher flow rates ($\gtrsim 40$~sccm), the titanium lifetime saturates at a maximum $\sim 6$~ms. We interpret this ``saturated'' value to be the ``pumpout'' time of the cell: on this timescale the titanium atoms are extracted through the aperture due to entrainment in the neon buffer gas \cite{hutzler2012buffer}.
		
We measured the beam of titanium atoms produced by the cell via spectroscopy after the cell. For measurements of atomic fluxes and transverse velocities, we  use absorption spectroscopy with the laser path transverse to the direction of the atomic beam. Typical powers are 30--40~$\mu$W and typical beam waists are $\sim$5--6~mm. Typical absorption signals show a main absorption pulse  preceded by 1--2 very temporally narrow ``pre-pulses''. The mechanism by which these pulses are produced is not known, but they are brief enough that few atoms are contained in them. As such, we focus only on the main pulse. This pulse first appears about 2--3~ms after ablation, with the peak occurring about 4--5~ms after ablation. As the transit time from the cell exit to the detection region is $\lesssim 350$~$\mu$s (as calculated from measured forward velocities), we attribute the delay primarily to the time it takes the titanium to be extracted from the cell. This delay is important for understanding the lack of signal seen for CH, as discussed in section \ref{sect:standard_cell}.
		
As previously reported in reference \cite{Nozzles2018}, the temperature of the titanium atoms in the beam increases as the flow rate is increased. Additionally, the temperatures also increase inside the cell with increasing flow rate, although after enough time they tend to approach similar values.
		
    \subsection{Small-volume cell} \label{subsec:SmallVolCellTiResults}

As expected, we see significantly shorter lifetimes for titanium in the small-volume cell. We expected the pumpout time to be proportional to cell volume and inversely proportional to the exit aperture area \cite{hutzler2012buffer}. This scaling would predict a pumpout time of $\sim$700~$\mu$s. Our measurements show even faster pumpout times. At early times after ablation, the titanium OD is not well described by an exponential. However, at later times, it appears roughly exponential, and at high neon buffer gas flow rates  ($\gtrsim 100$~SCCM), we measure a lifetime of 450$\pm$100~$\mu$s. We interpret this timescale as the pumpout time of the small-volume cell.
	
\begin{figure}[h!t]
    			\begin{center}
     			 \includegraphics[width=\linewidth]{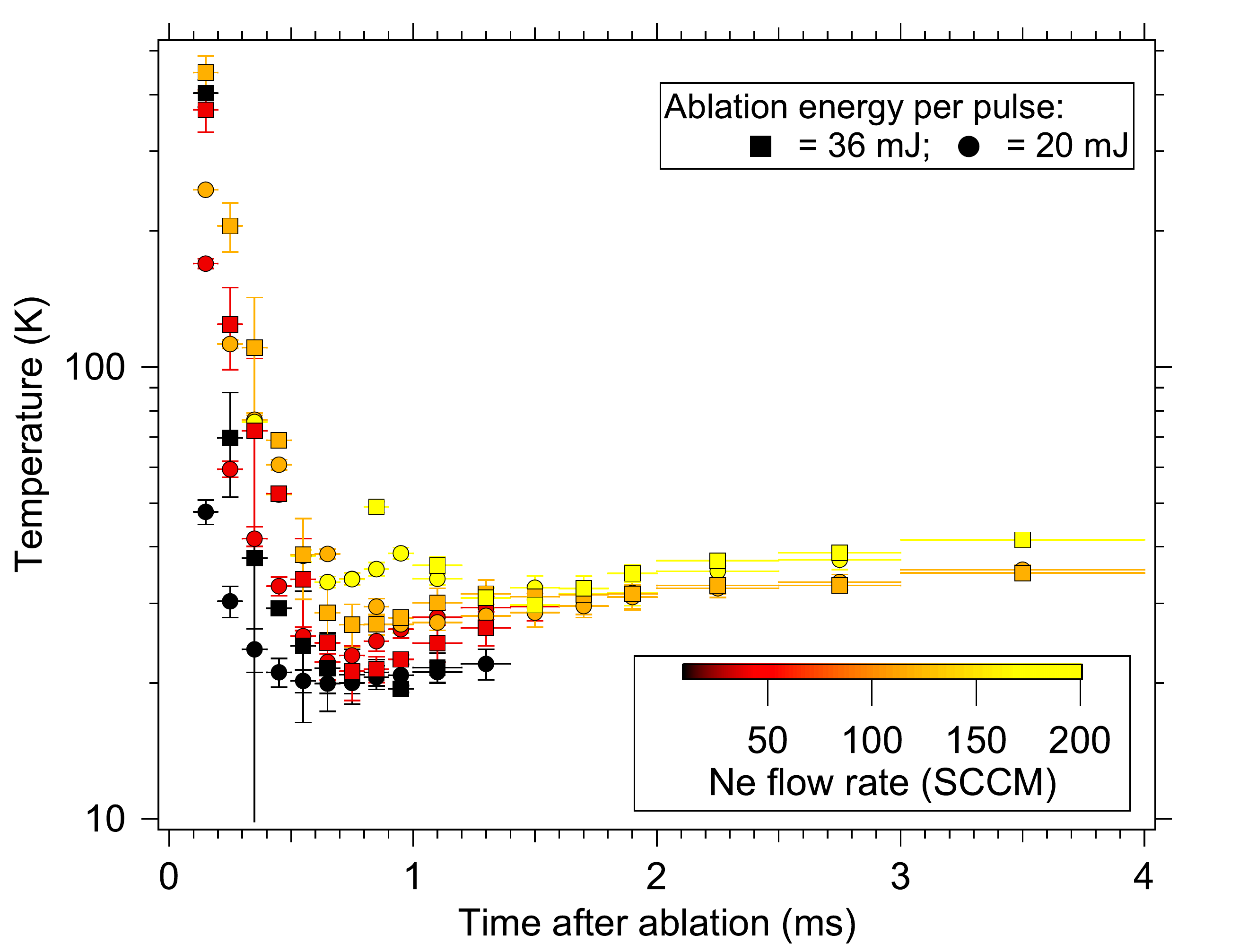}
    					\caption{
    						\label{fig:new_cell_Ti_temps_IC} 
    						Measured titanium translational temperatures (assuming linewidths are dominated by Doppler broadening) as a function of time. These were calculated from spectra taken at different time windows, with horizontal error bars showing the width of the window. The markers are placed over the middle of their respective time windows. The cell temperature was 16~K for this data.
					}     
    			\end{center}
		\end{figure}

Unfortunately, as seen in Fig. \ref{fig:new_cell_Ti_temps_IC}, the thermalization of the translational temperature of titanium occurs on a timescale comparable to the pumpout time. Consequently, we expect the atoms extracted at early times will produce a hotter beam.

We also note that at longer times after the ablation ($\gtrsim 1$~ms), the temperature of the titanium atoms in the cell appears to increase with time, as seen in Fig. \ref{fig:new_cell_Ti_temps_IC}. We are unsure as to what is the cause of this effect.

We measure the beam produced by the small-volume cell by the same methods outlined in Appendix \ref{subsec:StandardCellTiResults}.	Averaging over the  pulse of atoms (50--2300~$\mu$s after the ablation pulse), we find the titanium beam has a transverse velocity spread corresponding to a temperature of 46$\pm$10~K. The forward velocity of the titanium beam, as seen in Fig. \ref{fig:new_cell_Forward_velocities}, ranges from 200--260~m/s, increasing linearly with the flow rate. The axial velocity spread, shown in Fig. \ref{fig:new_cell_axial_temps}, shows a significant dependence on the buffer gas flow rate, exhibiting expansion cooling as the flow rate increases in the low rates, and then begins increasing after 40~SCCM. 
	
We observed a delay between ablation and the detection of Ti in the beam. Similar to the standard cell, we observe ``pre-pulses'' that occur before the primary measured pulse. Regarding the primary pulse, the delay time to initial signal is shorter, roughly 500~$\mu$s after ablation. The delay to the peak signal is $\sim1$~ms. Compensating for the delay due to the transit time from the cell aperture to the detection region, we find that the main pulse of Ti atoms first exits the cell $\sim 150~\mu$s after the ablation, with the peak occurring at $\sim 650~\mu$s. This reduction in the time it takes for atoms to leave the cell is comparable to the reduction in the lifetime observed in the cell.

	\section{Iodoform ablation} \label{sect:iodoformtgt}
	
As mentioned in section \ref{sect:300K}, we attempted to create CH ablation targets with iodoform powder. Initial attempts were done by melting iodoform in a beaker, resulting in a chalky rock that easily fractures. Ablating this resulted in the target quickly breaking apart while yielding no measurable quantity of CH. We also tried to recreate the iodoform target as described in \cite{fabrikant2014method}.
	
In that work, iodoform was dissolved in acetone and left in a glass beaker covered in aluminum foil at room temperature for roughly one month. At that time, the aluminum foil was found to be partially corroded, and the beaker contained both what appeared to be iodoform crystals as well as a black, ``gummy'' substance. In the prior work, ablation of the crystalline sample into a helium buffer gas in a cryogenic cell produced no measurable signal, but ablation of the black material produced large numbers of CH molecules with long lifetimes in the cell (although production was sometimes inconsistent shot-to-shot). Our recent attempts to produce the same material by this procedure failed, as did attempts to produce the material using heating. Typically, we created a black, hard substance with hints of crystalline structure. Ablating these materials did not yield measurable quantities of CH.

\end{document}